\documentclass[11pt]{article}
\usepackage{graphics}
\usepackage{epsfig}

\newcommand{\be}{\begin{eqnarray}}
\newcommand{\ee}{\end{eqnarray}}

\def\ll#1{\left#1}
\def\r#1{\right#1}
\def\fr{\frac{1}{2}}

\def\mref#1{(\ref{#1})}

\def\p{\partial}

\def\bd{\begin{displaymath}}
\def\ed{\end{displaymath}}

\def\bn{\begin{enumerate}}
\def\en{\end{enumerate}}
\def\ba#1{\begin{array}{#1}}
\def\ea{\end{array}}
\def\nn{\nonumber}
\newfont{\Bbb}{msbm10 scaled 1200}

\begin{document}

\pagestyle{empty}

\begin{center}

{\LARGE\bf Skeletons and superscars\\[0.5cm]}

\vskip 12pt

{\large {\bf Stefan Giller{$\dag$} and Jaros{\l}aw Janiak{$\ddag$}}}

\vskip 3pt

{$\dag$}Jan D{\l}ugosz University in Czestochowa\\
Institute of Physics\\
Armii Krajowej 13/15, 42-200 Czestochowa, Poland\\
e-mail: stefan.giller@ajd.czest.pl\\
$\ddag$Theoretical Physics Department II\\
University of {\L}\'od\'z,\\
Pomorska 149/153, 90-236 {\L}\'od\'z, Poland\\
e-mail: j.janiak@yahoo.pl

\end{center}
\vspace{6pt}

\begin{abstract}
Semiclassical wave functions in billiards based on the Maslov - Fedoriuk approach are constructed. They are
defined on classical constructions called skeletons which are the billiards generalization of Arnold's tori. Skeletons
in the rational polygon billiards considered in the phase space can be closed with a definite genus or can be open being a
cylinder-like or M{\"o}bius-like bands.
The skeleton formulation is applied to calculate semiclassical wave functions and the corresponding energy spectra for
the integrable and pseudointegrable billiards as well as in the limiting forms in some cases of chaotic ones. The superscars
of Bogomolny and Schmit are shown to be simply singular semiclassical solutions of the eigenvalue problem in the
billiards well built on the singular skeletons in the billiards with flat boundaries in
both the integrable and the pseudointegrable billiards as well as in the chaotic cases of such billiards.
\end{abstract}

\vskip 3pt
\begin{tabular}{l}
{\small PACS number(s): 03.65.-w, 03.65.Sq, 02.30.Jr, 02.30.Lt, 02.30.Mv} \\[1mm]
{\small Key Words: Schr{\"o}dinger equation, semiclassical expansion, Lagrange manifolds, classical}\\[1mm]
{\small trajectories, integrability, pseudointegrability, chaotic dynamics, quantum chaos, superscars}
\end{tabular}

\newpage

\pagestyle{plain}

\setcounter{page}{1}

\section{Introduction}

\hskip+2em Billiards while a non-analytic motion area are however well known as examples of the non-integrable two
dimensional
systems except the known cases of the integrable elliptical, rectangular and some triangle billiards. They are widely
considered
as a simple field of experimental \cite{13,27,14} as well as theoretical \cite{15,16,17,26} (and papers cited there)
and
computational investigations \cite{8,9,22} allowing to apply many different methods (see Sarnak's lecture
\cite{21}
and \cite{27} of the same author for an extensive review of the respective theoretical methods covering also billiards
manifolds as well as the students book of Tabachnikov \cite{29}).

In this paper we are going to
develop the semiclassical wave function (SWF) formalism which can be applied to non-integrable cases of the two dimensional
motions in billiards and which can be easily extended to higher dimensions.

Essentially our approach is very close to the one of Maslov and Fedoriuk \cite{4}. The main difference
between them is
in a treatment of crossing the singular points of the SWF's set on caustics. Namely, instead of making the canonical
phase
space variable transformations accompanied by the Fourier transformations of the SWF's to move through the caustic
points we apply the analytical continuation on the complex time plane to both the SWF's and the classical trajectories.
This greatly simplifies the corresponding procedure in comparison with the Maslov and Fedoriuk treatment and allows us
for not leaving the configuration space \cite{3}. However in this paper there is no opportunity to use the
simplification mentioned.

It is further a classical construction which we called {\bf skeletons} on which the SWF's are defined. Each skeleton is
compound in a closed way of bundles of rays (classical trajectories). SWF's of basic forms are defined just first on
bundles while the {\bf global} SWF's (GSWF) satisfying vanishing boundary conditions of Dirichlet or Neumann are sums of these
{\bf basic} SWF's (BSWF) and are uniquely and continuously defined on the skeleton. Skeletons play in this way a role
of Arnold's tori \cite{5} except that a number of ray bundles in skeletons can be infinite if billiards with chaotic motions
are considered.

When forms of billiards boundaries are considered and their relationships with types of the classical motion in them
one
realizes that billiards can be dividing into two general classes: the one with the flat boundaries,
i.e. the polygon billiards and the second class in which some pieces of their boundaries have finite
curvature.
In the polygon billiards there are no caustic phenomenon and this is the main and essential property
which differs both the classes.

However among the polygon billiards one can still distinguish a class of the rational billiards, i.e. which all angles
are rational part of $\pi$. As it was shown by Richens and  Berry \cite{8} (see also Tabachnikov \cite{29}) the rational
billiards can be classified
according to classical motions in the corresponding phase space which is performed on a set of two-dimensional disjoint compact
Lagrangian surfaces which collecting together can be made equivalent topologically to a compact close surface of a given genus
defined by the billiards. All possible values of genus can be realized by the rational billiards.

As an archetype of the rational pseudointegrable polygon billiards can be considered the broken rectangular billiards, i.e.
billiards which can be glued of a finite number of rectangles. The SWF's in the rectangular billiards and in the broken
ones are just studied intensively in this paper. Nevertheless other simple polygons such as the equilateral triangles and
the pentagon billiards are also considered.

It is well known that the semiclassical limits of wave functions are typically singular, i.e. the semiclassical wave functions
are frequently deprived of such properties as finiteness and smoothness which have to be satisfied by the exact wave functions.
Nevertheless SWF's alone as well as accompanied them energy spectra give typically very good approximations to the exact ones in the
high energy limit.

However it happens frequently that semiclassical calculations provide us with exact wave functions and energy spectra. This can
takes place for example when the calculated SWF's satisfy the same conditions as the exact ones and the classically
allowed configuration space coincides with the quantum one. Such SWF's are called {\bf regular} in this paper. Examples of such
regular semiclassical solutions are provided by the generic SWF in the rectangle, the equilateral triangle and the pentagon billiards.

Most of the semiclassical wave functions which solve the eigenvalue problem in the semiclassical limit are however
{\bf singular}, i.e. they
do not satisfy some of the conditions which the exact solutions have to do. The SWF's which
can be constructed for the rectangular billiards and the broken
rectangle ones are not exceptional in the respects mentioned and can be also of two
types - the regular and the singular ones depending on skeletons used to their constructions. It is just the singular
SWF's for which superscars phenomenon of Bogomolny {\it et al} \cite{1,2} can be observed. While this singular behaviour of SWF's in the
cases considered is reduced only to discontinuities of the first derivatives of the respective SWF's this "defect" of
them is enough for being a source of the superscar phenomenon.

It is also shown in the paper that the superscars are typical not only for the integrable and pseudointegrable systems
but also for the chaotic ones. A well known example is the Bunimovich stadium with its bouncing ball modes \cite{14,15,16}.
But it is
easy to give many other examples of the chaotic billiards with any form of the superscars which can be found in the
broken rectangular billiards.

The paper is organized as follows.

In the next section the Maslov - Fedoriuk method of the semiclassical wave function construction is reminded and
discussed.

In sec.3 a construction of skeletons is given.

In sec.4 global SWF's in billiards are constructed.

In sec.5 the rectangular billiards, the broken rectangle ones, the triangle and the pentagon billiards are considered.
In the rectangular billiards case all possible SWF's which can be defined in it, i.e. the regular and the singular ones,
are discussed.
SWF's in the remaining billiards are considered selectively because of increasing complexity of the corresponding
skeletons. Nevertheless it is shown that in these cases the superscar SWF's are common and more
spectacular than for the rectangular billiards.

Next in sec.6 it is shown that the superscar SWF's can be implemented into
chaotic deformations of the broken rectangle, the triangle and the pentagon billiards.

In sec.7 the results of the paper are summarized.

There are two appendixes attached to the paper which justify the main assumptions used in the construction of the
global SWF's on skeletons.

\section{Semiclassical wave function expansion for $n$-D stationary Schr{\"o}dinger equation}

\hskip+2em Consider the $n$-dimensional stationary Schr{\"o}dinger equation:
\be
\bigtriangleup\Psi({\bf r})+\lambda^2\frac{2m}{\hbar^2}(E-V({\bf r}))\Psi({\bf r})=0
\label{1}
\ee
with a potential $V({\bf r}),\;{\bf r}\in R_n$ confining a point particle with a mass $m$ and containing a formal dimensionless parameter
$\lambda>0$. For a convenience we shall put further $\hbar=1$ and $m=1$. The Schr{\"o}dinger equation is recovered by
putting $\lambda=1$ in \mref{1}.

We would like to construct a solution to Eq.\mref{1} using the idea of Maslov {\it et al} \cite{4} and considering the wave function $\Psi({\bf r})$
as defined on families of classical trajectories a dynamic of which is given by the classical Hamiltonian $H=\fr{\bf p}^2+V({\bf r})$ and which
carry an energy $E_0$ all.

Such families are constructed locally in the following way.

In $R_n$ we choose a $n-1$-D hypersurface $\Sigma_{n-1}$ parametrized by local coordinates ($s_1,...,s_{n-1}$). On
$\Sigma_{n-1}$ the initial momenta
${\bf p}({\bf r}_0),\;{\bf r}_0\in\Sigma_{n-1}$, are defined so that the pair
$({\bf r}_0,{\bf p}({\bf r}_0)),\;{\bf r}_0\in\Sigma_{n-1}$ serve as the initial data for the trajectories
${\bf r}(t)={\bf f}({\bf r}_0,{\bf p}({\bf r}_0);t)$ developed by the Hamiltonian $H$. Additionally the momentum field
${\bf p}({\bf r}_0)$ defined on $\Sigma_{n-1}$ has to satisfy:
\be
\oint_C{\bf p}({\bf r}_0)d{\bf r}_0=0
\label{1a}
\ee
for each loop $C,\;C\subset \Sigma_{n-1}$.

We can now define the transformation: ${\bf r}\to(t,{\bf r}_0)\to(t,s_1,...,s_{n-1}),
\;{\bf r}\equiv(x_1,...,x_n),\;{\bf r}_0\equiv(x_{0,1}(s_1,...,s_{n-1}),...,x_{0,n}(s_1,...,s_{n-1}))$,
which is one-to-one up to a caustic surface $C_{n-1}$ on which the
Jacobean (${\bf f}({\bf r}_0,{\bf p}({\bf r}_0);t)\equiv{\bar{\bf f}}(t,s_1,...,s_{n-1})$):
\be
J(t,s_1,...,s_{n-1})=\ll|\frac{\p {\bar f}_i}{\p t},\frac{\p {\bar f}_i}{\p s_j}\r|
\label{4}
\ee
vanishes.

A $n$-dimensional domain $\Lambda_n$ of $2n$-dimensional phase space $R_{2n}$ made in this way by the hypersurface
$\Sigma_{n-1}$ and trajectories emerging from it is known as the Lagrange manifold \cite{5}.

Therefore in the variables $t,s_1,...,s_{n-1}$ the new wave function $\chi(t,s_1,...,s_{n-1})$ satisfies the following relation with the previous one:
\be
|\chi(t,s_1,...,s_{n-1})|^2=|\Psi({\bar{\bf f}}(t,s_1,...,s_{n-1})|^2|J(t,s_1,...,s_{n-1})|
\label{4a}
\ee

The particle momentum ${\bf p}$ on the trajectories ${\bf r}(t)={\bar{\bf f}}(t,s_1,...,s_{n-1})$ satisfies of course the equation:
\be
\frac{\p{\bar{\bf f}}(t,s_1,...,s_{n-1})}{\p t}={\bf p}({\bar{\bf f}}(t,s_1,...,s_{n-1}))
\label{5}
\ee
defining also the Jacobean evolution. Namely:
\be
\frac{\p}{\p t}\frac{\p{\bar{f}}_i(t,s_1,...,s_{n-1})}{\p t}=\sum_{j=1}^n
\frac{\p p_i}{\p x_j}\frac{\p{\bar{f}}_j(t,s_1,...,s_{n-1})}{\p t}\nn\\
\frac{\p}{\p t}\frac{\p{\bar{f}}_i(t,s_1,...,s_{n-1})}{\p s_l}=\sum_{j=1}^n
\frac{\p p_i}{\p x_j}\frac{\p{\bar{f}}_j(t,s_1,...,s_{n-1})}{\p s_l}\nn\\l=1,...,n-1
\label{6}
\ee

so that
\be
\frac{\p J(t,s_1,...,s_{n-1})}{\p t}=J(t,s_1,...,s_{n-1})\nabla {\bf p}({\bar{\bf f}}(t,s_1,...,s_{n-1}))
\label{7}
\ee

The above equation is just the Liouville theorem with the solution:
\be
J(t,s_1,...,s_{n-1})=J(s_1,...,s_{n-1})e^{\int_0^t\nabla {\bf p}({\bar{\bf f}}(t',s_1,...,s_{n-1}))dt'}
\label{7a}
\ee
where $J(s_1,...,s_{n-1})$ is the value of the Jacobean on the hypersurface $\Sigma_{n-1}$.

It is well known from the classical Hamiltonian mechanics \cite{5} that the action integral:
\be
S({\bf r},{\bf r}_0)=\int_{{\bf r}_0}^{\bf r}{\bf p}({\bf r}')d{\bf r}'
\label{7b}
\ee
taken on the Lagrange manifold $\Lambda_n$ is a point function of ${\bf r}$ and ${\bf r}_0$. Therefore taking ${\bf r}_0$ as a definite fixed
point of the hypersurface $\Sigma_{n-1}$ and denoting by $S({\bf r})$ the action function corresponding to this case we can complete
a definition of the wave function $\chi(t,s_1,...,s_{n-1})$ by the following equation:
\be
\Psi^\sigma({\bar{\bf f}}(t,s_1,...,s_{n-1}))=J^{-\fr}(t,s_1,...,s_{n-1}))
e^{\sigma\lambda iS({\bar{\bf f}}(t,s_1,...,s_{n-1})}\chi^\sigma(t,s_1,...,s_{n-1})
\label{2}
\ee
where $\sigma=\pm$ is a signature of $\Psi^\sigma({\bf r})$.

The form \mref{2} of the semiclassical wave functions (SWF) will be called {\bf basic} (BSWF).

Therefore the quantities involved in the above definitions satisfy the following equations:
\be
{\bf p}({\bf r})=\nabla S({\bf r})\nn\\
\fr{\bf p}^2({\bf r})+V({\bf r})-E_0=0\nn\\
\triangle(J^{-\fr}\chi^\sigma({\bf r}))+\sigma 2i\lambda J^{-\fr}({\bf r})
\nabla \chi^\sigma({\bf r})\cdot{\bf p}({\bf r})+2\lambda^2(E-E_0)J^{-\fr}({\bf r})\chi^\sigma({\bf r})=0\nn\\
{\bf r}={\bar{\bf f}}(t,s_1,...,s_{n-1})
\label{3}
\ee

By the variables $t,s_1,...,s_{n-1}$ the third of the last equations can be rewritten in the following form:
\be
\sigma 2i\lambda\frac{\p\chi^\sigma(t,s_1,...,s_{n-1},\lambda)}{\p t}+\nn\\
J^{\fr}\triangle\ll(J^{-\fr}\chi^\sigma(t,s_1,...,s_{n-1},\lambda)\r)+
\lambda^2(E-E_0)\chi^\sigma(t,s_1,...,s_{n-1},\lambda)=0
\label{11}
\ee
where a dependence of $\chi^\sigma(t,s_1,...,s_{n-1},\lambda)$ on $\lambda$ was shown explicitly.

The Eq.\mref{11} describes the time evolution of $\chi^\sigma(t,s_1,...,s_{n-1},\lambda)$ along trajectories starting
on the hypersurface $\Sigma_{n-1}$ if its "initial" values on this surface, i.e.
$\chi^\sigma(0,s_1,...,s_{n-1},\lambda)\equiv\chi^\sigma(s_1,...,s_{n-1},\lambda)$ are given.

We are going to consider the equation \mref{11} in the semiclassical limit $\lambda\to+\infty$ looking for its solutions in
the form of the following asymptotic series:
\be
E-E_0=\sum_{k\geq 1}E_k\lambda^{-k-1}\nn\\
\chi^\sigma(t,s_1,...,s_{n-1},\lambda)=\sum_{k\geq 0}\chi_k^\sigma(t,s_1,...,s_{n-1})\lambda^{-k}\nn\\
\chi^\sigma(s_1,...,s_{n-1},\lambda)=\sum_{k\geq 0}\chi_k^\sigma(s_1,...,s_{n-1})\lambda^{-k}
\label{3a}
\ee

Putting $\lambda=1$ in \mref{1}, \mref{2} and \mref{3a} we get approximate semiclassical solutions to the energy
eigenvalue problem of the Schr{\"o}dinger equation.

It is to be noticed that for the selfconsistency reasons the semiclassical series for the energy parameter in
\mref{3a} starts from
the second power of $\lambda^{-1}$, i.e. this ensures the proper hierarchy of steps in the algorithm of semiclassical
calculations by which the higher order terms of the series in \mref{3a} are determined by the lower order ones.

It should be noticed also that despite the fact that $E_0$ enters the classical equation of motion \mref{3} it is still
quantum, i.e. its value depends on $\hbar$ which is considered to have the definite numerical value, i.e. $\hbar$ is not
a parameter. In particular the series \mref{3a} represent the inverse power hierarchy in the formal parameter $\lambda$,
i.e. not in powers of $\hbar$, between subsequent terms.

Moreover $E_0$ if quantized can depend on $\lambda$. However, whatever this dependence is the semiclassical series of
the difference $E-E_0$ must be given by \mref{3a}.

Needless to say the introducing $\lambda$ makes a treatment of the Schr{\"o}dinger equation equivalent of course
to considering it in the limit $\hbar\to 0$, i.e.
semiclassically, clearly however separating the role of $\hbar$ as a parameter from its role defining the microscale
of quantum phenomena.

Substituting \mref{3a} into \mref{11} we get:
\be
\frac{\p\chi_0^\sigma(t,s_1,...,s_{n-1})}{\p t}=0\nn\\
\frac{\p\chi_{k+1}^\sigma(t,s_1,...,s_{n-1})}{\p t}=\nn\\
\frac{\sigma i}{2}\ll(J^{\fr}\triangle\ll(J^{-\fr}\chi_{k}^\sigma(t,s_1,...,s_{n-1})\r)+
2\sum_{l=0}^kE_{k-l+1}\chi_l^\sigma(t,s_1,...,s_{n-1})\r)\nn\\
k=0,1,2,...,
\label{12}
\ee
with the obvious solutions:
\be
\chi_0^\sigma(t,s_1,...,s_{n-1})\equiv\chi_0^\sigma(s_1,...,s_{n-1})\nn\\
\chi_{k+1}^\sigma(t,s_1,...,s_{n-1})=\chi_{k+1}^\sigma(s_1,...,s_{n-1})+\nn\\
\frac{\sigma i}{2}\int_0^t\ll(J^{\fr}\triangle\ll(J^{-\fr}\chi_{k}^\sigma(t',s_1,...,s_{n-1})\r)+
2\sum_{l=0}^kE_{k-l+1}\chi_l^\sigma(t',s_1,...,s_{n-1})\r)dt'\nn\\
k=0,1,2,...,
\label{12z}
\ee

For future applications it is worth to note that if \mref{11} is obviously invariant on a reparametri\-zation of the hypersurface $\Sigma_{n-1}$
it is also invariant on the following change of variables:
\be
t\to \tau(s_1,...,s_{n-1})\pm t\nn\\
s_k\to h_k(s_1,...,s_{n-1})\nn\\
k=1,...,n-1
\label{13a}
\ee
if it is accompanied simultaneously by the transformations:
\be
\chi^\sigma(t,s_1,...,s_{n-1},\lambda)\to\nn\\{\chi}^{\pm\sigma}(t,s_1,...,s_{n-1},\lambda)\equiv
(\pm J_1)^{-\fr}(h_1^{-1}(s_1,...,s_{n-1}),...,h_{n-1}^{-1}(s_1,...,s_{n-1}))\times\nn\\
\chi^\sigma(\tau(s_1,...,s_{n-1})\pm t,h_1^{-1}(s_1,...,s_{n-1}),...,h_{n-1}^{-1}(s_1,...,s_{n-1}),\lambda)
\label{13b}
\ee
where $J_1(s_1,...,s_{n-1})$ is the Jacobean of the transformation $s_k\to h_k(s_1,...,s_{n-1}),\;k=1,...,n-1$.

\section{Skeletons - classical constructions in billiards}

\hspace{15pt}Before applying the above formalism to construct continuous semiclassical wave functions inside billiards
$B$ (Fig.1) vanishing on its boundary i.e. satisfying the Dirichlet boundary conditions it is first necessary to
perform a classical construction consisting of classical trajectories on which
the desired SWF's can be defined. This is just {\bf a skeleton} i.e. a closed set of families of trajectories which
forms a base on which SWF's can be constructed.

The skeleton idea relies on an observation that the short wave packets propagate in billiards approximately along the
straight lines (rays of the geometrical optics) gradually however becoming wider and wider due to unavoidable diffractive effects.
Because of the last effects even an initially narrow bundle of such rays fills completely a volume admitted by the
respective boundary conditions.

Therefore from the one hand the SWF propagates along the classical objects - trajectories,
but from the other hand it has to be defined on a bundle of such trajectories sufficiently wide to gather diffractive
effects satisfying nevertheless the rules of the geometrical optics. In particular an effect of shadow typical for the
geometrical optics should be observed on a way of propagation of SWF's.

Finally a set of all such bundles should be closed and mutually connected to permit us performing a construction
of the global semiclassical wave function from its pieces defined on the separate bundles.

Below a notion of a skeleton is defined which according to our expectations takes into account all the aspects of the
semiclassical limit discussed above.

\subsection{Ray bundles and bundle skeletons}

\hspace{15pt}For the needs of this paper we shall assume that the billiards $B$ is classical and according to Fig.1 it has no holes inside and
its boundary $\p B$ is a
\begin{figure}
\begin{center}
\psfig{figure=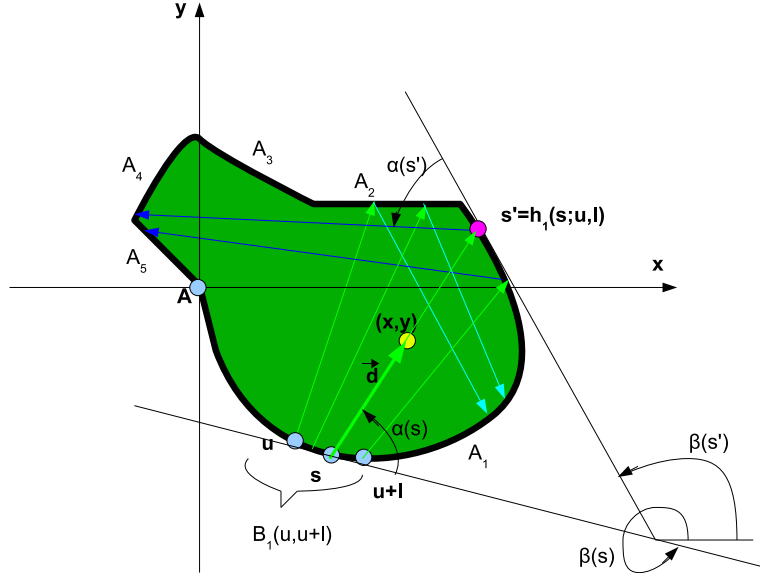,width=12cm}
\caption{An arbitrary billiards}
\end{center}
\end{figure}
closed curve independent of $\lambda$ and given by
${\bf r}={\bf r}_0(s)=[x_0(s),y_0(s)]$ where $s$
is a distance of a boundary point ${\bf r}_0(s)$ measured clockwise along $\p B$ from some other point $A$ of $\p B$ chosen arbitrary, i.e.
$s(A)=0$. Both $x_0(s)$ and $y_0(s)$ are continues. The curve however consists
of a finite number $q,\;q\geq 1$ of smooth arcs or segments $A_1,...,A_q$ with respective length $L_1,...,L_q$, so that the
derivatives $x_0'(s)$ and $y_0'(s)$ are discontinuous
in a finite number of points on the segment $0\leq s\leq L$ where $L=L_1+\cdots+L_q$ is the global length of $\p B$. Both
$x_0(s)$ and $y_0(s)$ are of course periodic with the period equal to $L$. We shall identify the point $A$ with the
point beginning the arc (segment) $A_1$.

Next we define a bundle of rays as a family of trajectories in the following way.

Let $A_k(u,l),\;L_1+\cdots+L_{k-1}\leq u\leq L_1+\cdots+L_k,\;0<l\leq L_k,$ be an open connected piece of the
arc (segment) $A_k$ beginning at $s=u$ and having a length $l$.

Let further ${\bf r}_k(s,t;u,l),\;u<s<u+l,\;0\leq t,$ be a family of trajectories given by
angles $\gamma_k(s;u,l),\;0\leq\gamma_k(s;u,l)\leq 2\pi$, at which the trajectories escape from  $A_k(u,l)$. The angles
$\gamma(s;u,l)$ are smooth functions of $s$ and are measured with respect to the $x$-axis while the tangential vectors
${\bf t}(s)=[\frac{dx_0(s)}{ds},\frac{dy_0(s)}{ds}]=[\cos\beta(s),\sin\beta(s)]$ are inclined to the $x$-axis at
angles $\beta(s)$ (Fig.1). The latter angle can be discontinuous at the points where $x_0'(s)$ and $y_0'(s)$ are
discontinuous. Then the angle $\alpha_k(s;u,l)=\gamma_k(s;u,l)-\beta_k(s)$ is made by
the classical ball momentum ${\bf p}(s;u,l)$ on the trajectory with the tangent vector ${\bf t}(s)$, i.e.
${\bf p}(s;u,l)\cdot{\bf t}(s)=p\cos\alpha_k(s;u,l)$. It is assumed that $0<\alpha_k(s;u,l)<\pi$.

The classical time evolution of the family ${\bf r}_k(s,t;u,l),\;u<s<u+l,$ is therefore the following
\be
{\bf r}_k(s,t;u,l)={\bf r}_0(s)+{\bf p}(s;u,l)t,\;\;\;\;\;\;{\bf r}_0(s)\in A_k
\label{15}
\ee
where ${\bf p}(s;u,l)=[p\cos\gamma_k(s;u,l),p\sin\gamma_k(s;u,l)]$ satisfies the classical equations of motions
\mref{3}, i.e. ${\bf p}^2(s;u,l)=2E_0$ (again we put $m=1$ for the billiard ball mass).

The trajectories \mref{15} define of course the change of variables $(x,y)\to(t,s)$, in vicinity of $A_k(u,l)$, i.e.
$x=f_k(t,s;u,l),\;y=g_k(t,s;u,l)$ with the Jacobean:
\be
{\tilde J}_k(t,s;u,l)=p^2\gamma_k'(s;u,l)t-p\ll|{\bf t}(s)\r|\sin\alpha_k(s;u,l)=
p^2\gamma_k'(s;u,l)t-p\sin\alpha_k(s;u,l)\nn\\
\label{16}
\ee
since $\ll|{\bf t}(s)\r|=1$.

The family of trajectories defined in the above way will be called a {\bf bundle of rays} emerging from the segment
$A_k(u,l)$ of $A_k$ and will be denoted by $B_k(u,l)$ while the trajectories themselves will be called {\bf rays}.

Note
that each bundle by definition is an open set of rays with at most two limiting rays as its boundary. If the limiting
rays coincide then necessarily the corresponding segment $A_k(u,l)$ can be closed to the whole billiards boundary.
This is the case for example of a bundle defined in the circle billiards which rays have the same angular momenta each.

Suppose the limiting rays of bundles $B_k(u,l)$ and $B_{k'}(u',l')$ coincide on a piece $P$ of them. Then by closing the
bundles on this common boundary we get a bundle which we call {\bf compound} and
denote by $B_k(u,l)\cup_P\ B_{k'}(u',l')$. Such a closing operation will be called a {\bf composition} of the initial
bundles. Of course each bundle can be decomposed into two others by the reverse operation becoming the composition of
the resulting bundles.

A compound bundle can be composed of course from many bundles and its rays emerge then from the sum $\bigcup_kA_k(u,l)$
of the segments of the composing bundles. If $\bigcup_kA_k(u,l)$ is a connected piece of the billiards boundary then
such a composition and its result a compound bundle will be called {\bf regular}. Other cases of compositions and
compound bundles will be called {\bf singular}.

Since each ray of the bundle $B_k(u,l)$ after some time $\tau_k(s;u,l),\;{\bf r}_0(s)\in A_k(u,l)$, (different for
different rays)
achieves another point of the boundary $\p B$ it means that the bundle $B_k(u,l)$ maps the segment $A_k(u,l)$ into
another piece $BA_k(u,l)$ of the boundary $\p B$. In general this map of $A_k(u,l)$ into
$BA_k(u,l)$ provided
by the transformation \mref{15} is one-to-one except the caustic points of $BA_k(u,l)$ in which
${\tilde J}_K(\tau_k(s;u,l),h_k(s;u,l);u,l)=0,\;{\bf r}_0(h_k(s;u,l))\in BA_k(u,l)$. Here $h_k(s;u,l),\;
{\bf r}_0(s)\in A_k(u,l)$, realizes explicitly this map. If however $\frac{\p h_k(s;u,l)}{\p s}
\neq 0,\;{\bf r}_0(s)\in A_k(u,l)$, i.e. this map is one-to-one then such a bundle will be called regular
at the boundary $\p B$ or simply regular.

All bundles considered below will be assumed to be regular.

By $DB_k(u,l)$ will be denoted a domain of the billiards $B$ covered by rays of the bundle $B_k(u,l)$ emerging from
$A_k(u,l)$ and ending at $BA_k(u,l)$. The domain $DB_k(u,l)$ is locally a Lagrangian manifold on which each loop
integral $\oint{\bf p}\cdot d{\bf r}$ vanishes.

While a definition of a bundle is essentially local it can happen that in particular cases of bundles or
compound bundles they can cover the whole billiards except necessarily some pieces of its boundary, i.e.
$\overline{DB_k(u,l)}=B=\overline{D(B_k(u,l)\cup_P\ B_{k'}(u',l'))}$. Such bundles or compound bundles will be then
called {\bf global}.

Assume further the boundary $\p B$ to be a mirror-like, i.e. reflecting the incoming rays according to the reflection
principle of the geometrical optics and let $A_{k'}(u',l')$ be a piece of another arc $A_{k'}$ of $\p B$ such that
$A_{k'}(u',l')\cap BA_k(u,l)\neq\oslash$ on which another regular bundle of rays
$B_{k'}(u',l')=\{{\bf r}_{k'}(s,t;u',l'):u'<s<u'+l',\;0\leq t\}$ is defined.

If on the segment $A_{k'}(u',l')\cap BA_{k}(u,l)$ the ray bundle $B_{k'}(u',l')$ coincides with the reflected one we
call the ray bundle $B_{k'}(u',l')$ a reflection of the bundle $B_k(u,l)$ on the segment mentioned.

The reflection operation over the bundle $B_k(u,l)$ will be denoted by $\Pi$ so that $\Pi B_k(u,l)$ denotes the set of
all rays arising by the reflection of the bundle $B_k(u,l)$ of $\p B$.

Consider now a family of the disjoint ray bundles ${\bf B}=\bigcup B_k(u,l),\;B_k(u,l)\cap B_{k'}(u',l')=
\oslash$ if $B_k(u,l)\neq B_{k'}(u',l')$.

The family {\bf B} will be called {\bf closed} under reflection $\Pi$ on the boundary $\p B$ if the following two conditions
are satisfied for each bundle $B_k(u,l)\in{\bf B}$:
\be
\Pi B_k(u,l)=\bigcup_{j=1}^n B_j(u_j,l_j)\cap \Pi B_k(u,l),\;\;\;\;\;\;\;B_j(u_j,l_j)\in A_{i_j},\;j=1,...,n\nn\\
B_k(u,l)=\bigcup_{i=1}^m \Pi B_i(u_i,l_i)\cap B_k(u,l),\;\;\;\;\;\;\;B_i(u_i,l_i)\in A_{j_i},\;i=1,...,m\nn\\
\Pi{\bf B}={\bf B}
\label{16a}
\ee

A closed family ${\bf B}'=\bigcup B_{k'}(u',l')$ is embedded into a closed family ${\bf B}=\bigcup B_k(u,l)$ if each ray
bundle of ${\bf B}'$ is a subset of some ray bundle of ${\bf B}$ and each bundle of ${\bf B}$ contains some bundle of
${\bf B}'$.

A closed bundle family is called connected if a unique possibility to represent it by a sum of another two disjoint
closed bundle families is a decomposition operation done on every bundle of the family.

A closed connected bundle family will be called a {\bf Lagrange bundle skeleton} or simply a {\bf skeleton} if it cannot
be embedded into another closed connected bundle family.

From now on all considered bundle families will be assumed to be skeletons.

Let us stress the following four basic properties of skeletons which are of great importance:
\begin{enumerate}
\item each skeleton is complete i.e. none additional bundle can be added to it not destroying it connectedness,
\item each skeleton cannot be decomposed into "smaller" ones (again by the connectedness property),
\item each bundle of a given skeleton is a result of reflections of other bundles of the same skeleton (by \mref{16a}),
\item each ray belonging to a skeleton ${\bf B}$ will never leave ${\bf B}$ by its time evolution and bounces on
the billiards boundary.
\end{enumerate}

Let us now reverse in time all trajectories belonging to ${\bf B}$. This operation leads us again to some skeleton
${\bf B}^A$ which will be called {\bf associated} with ${\bf B}$.

Bundles of ${\bf B}^A$ are obtained simply from the corresponding bundles
of ${\bf B}$. Namely with each bundle $B_k(u,l)$ of ${\bf B}$ let us associate a bundle $B_k^A(u,l)$ which
trajectories satisfy the following condition:
\be
\gamma_k^A(s;u,l)=\pi+2\beta(s)-\gamma_k(s;u,l),\;\;\;\;u<s<u+l
\label{16b}
\ee
i.e. these trajectories are just the reflections on $\p B$ of the time reversed trajectories defined by
$\gamma_k(s;u,l)$ and belonging to $B_k(u,l)$.

The skeleton ${\bf B}^A$ is organized by all bundles $B_k^A(u,l)$ obtained in the above way.

Needless to say $({\bf B}^A)^A\equiv {\bf B}$.

Finally let us denote by $D_{\bf B}\subset B$ a topological sum of all domains $DB_k(u,l)$, i.e. $D_{\bf B}=
\bigcup_{B_k(u,l)\subset{\bf B}}DB_k(u,l)$. Define $D_{{\bf B}^A}$ analogously. By the construction of
the skeleton ${\bf B}^A$ we have $D_{{\bf B}^A} \equiv D_{\bf B}$.

An useful operation on a skeleton is its reduction which means making all possible compound bundles of the bundles
of the skeleton. Such a form of the skeleton will be called a {\bf reduced skeleton} and denoted by ${\bf B}^R$.

Reduced skeletons although useful are deprived however of many properties of the skeletons. In particular it is typically
not possible to construct a skeleton associated with the reduced one.  On the other hand in the polygon billiards
each compound bundle is associated with a definite momentum of the billiards ball so that to different momenta of the
ball correspond different compound bundles.

If a reduced skeleton ${\bf B}^R$ contains only global bundles then the skeleton {\bf B} will be called
{\bf global} and if its bundles are all global and regular the skeleton {\bf B} will be called {\bf regular}.
Skeletons which are not regular will be called {\bf singular}. It then follows that skeletons can be global but singular.

\subsection{Skeletons in the phase space. Pseudointegrable billiards}

\hspace{15pt}In the case of the polygon billiards skeletons considered in the phase space are decomposed into separate
pieces parallel to the billiards
plane. Each a piece corresponds to a separate compound bundle. Any trajectory of the skeleton visits all pieces (bundles)
jumping from one piece (bundle) to another when achieving the piece boundary which projected on the billiards plane
coincides with the corresponding piece of the billiards boundary.

One can imagine a continues surface made of the pieces mentioned gluing their respective boundaries.
Namely, we can get it by gluing points of a piece boundary by which trajectories leave the piece to visit the next
one with these
entry points of the next piece. If a number of bundles is finite such a construction leads us to a compact two
dimensional surface. If such a surface is closed it has then a definite genus. Among the others these are the cases of
the global skeletons in the rational polygon billiards \cite{8,29,6}.

It will appear in sec.5 that the skeletons in the investigated billiards can form Lagrange surfaces in the phase space
which are closed according to the above construction with a definite genus but also which are open being a
cylinder-like or a M{\"o}bius-like bands.

It will appear also in the next sections that only the skeletons which are regular can provide us with GSWF's which are exact
solutions of the eigenvalue problems. The singular skeletons provides us with singular GSWF's which cannot be exact and
some of them show typical properties of the superscar solutions.

Billiards for which {\bf every} possible skeleton has a finite number of bundles are certainly
distinguished by
a possibility of construction of all GSWF's for such a billiards in compact finite forms. Therefore it would be reasonable to
extend a notion of pseudointegrability introduced by Richens and Berry \cite{8} to such a billiards despite the fact
that not all of the skeletons corresponding to them can be global so that the continuous Lagrange surfaces which
construction has been described above can be open, i.e. with boundaries.

\section{SWF's defined on a skeleton}

\subsection{BSWF's defined on a bundle}

\hspace{15pt}Consider a skeleton ${\bf B}$. On each of its ray bundle $B_k(u,l)$ we can now define the following
pair of BSWF's $\Psi_k^\sigma(t,s;u,l;\lambda),\;\sigma=\pm$:
\be
\Psi_k^\sigma(t,s;u,l;\lambda)={\tilde J}_k^{-\fr}(t,s;u,l)
e^{\sigma i\lambda\ll(p^2t+p\int_u^s\cos\alpha_k(s;u,l)ds'\r)}\chi_k^\sigma(t,s;u,l;\lambda)
\label{17}
\ee
where $p^2=2E_0$ and $\chi_k^\sigma(t,s;u,l;\lambda),\;\sigma=\pm$, are given by \mref{3a}
and \mref{12z}.

Exactly in the same way we can define a pair $\Psi_{A;k}^\sigma(t,s;u,l;\lambda),\;\sigma=\pm$, of BSWF's on the
corresponding associated bundle $B_k^A(u,l)$:
\be
\Psi_{A;k}^\sigma(t,s;u,l;\lambda)={\tilde J}_{A;k}^{-\fr}(t,s;u,l)
e^{\sigma i\lambda\ll(p^2t-p\int_u^s\cos\alpha_k(s;u,l)ds'\r)}\chi_{A;k}^\sigma(t,s;u,l;\lambda)
\label{17a}
\ee

It will be also convenient for further considerations to substitute the time variable $t$ by the distance variable
$d=pt$ and consequently to give the trajectories \mref{15}, the Jacobean \mref{16}
and the wave function \mref{17} the following forms:
\be
{\bf r}_k(d,s;u,l)={\bf r}_0(s)+{\bf d}(s;u,l)\nn\\
{\bf d}(s;u,l)={\bf p}t=[d\cos\gamma_k(s;u,l),d\sin\gamma_k(s;u,l)]\nn\\
{\bf r}_0(s)\in A_k(u,l)
\label{15a}
\ee
and
\be
J_k(d,s;u,l)=\frac{1}{p}{\tilde J}_k(t,s;u,l)=\frac{\p\gamma_k(s;u,l)}{\p s}d-\sin\alpha_k(s;u,l)\nn\\
{\bf r}_0(s)\in A_k(u,l)
\label{16c}
\ee
and
\be
\Psi_k^\sigma(d,s;u,l;\lambda)=J_k^{-\fr}(d,s;u,l)
e^{\sigma i\lambda p\ll(d+\int_u^s\cos\alpha_k(s';u,l)ds'\r)}{\bar\chi}_k^\sigma(d,s;u,l;\lambda)\nn\\
{\bf r}_0(s)\in A_k(u,l)
\label{17A}
\ee
where ${\bar\chi}_k^\sigma(d,s;u,l;\lambda)\equiv p\chi_k^\sigma(\frac{d}{p},s;u,l;\lambda)$ and $\sigma=\pm$.

Nevertheless, for simplicity of notations, the bar over ${\bar\chi}_k^\sigma(d,s;u,l;\lambda)$ will be dropped in our
further considerations.

By the variable $d$ the solutions \mref{12z} can be rewritten in the form:
\be
{\chi}_{k,0}^\sigma(d,s;u,l)\equiv{\chi}_{k,0}^\sigma(s;u,l)\nn\\
{\chi}_{k,j+1}^\sigma(d,s;u,l)={\chi}_{k,j+1}^\sigma(s;u,l)+\nn\\
\frac{\sigma i}{2p}\int_0^d\ll({\tilde\triangle}_k(a,s;u,l){\chi}_{k,j}^\sigma(a,s;u,l)+
2\sum_{m=0}^jE_{j-m+1}{\chi}_{k,m}^\sigma(a,s;u,l)\r)da\nn\\
j=0,1,2,...,
\label{13A}
\ee
where ${\tilde\triangle}_k(d,s;u,l)=J_k^\fr(d,s;u,l)\cdot\triangle_k(d,s;u,l)\cdot J_k^{-\fr}(d,s;u,l)$ and $\triangle_k(d,s;u,l)$ is
the Laplacean expressed by the variables $d$ and $s$ corresponding to the -bundle.

$\Psi_k^\sigma(d,s;u,l;\lambda),\;\sigma=\pm,$ are defined initially in the domain $D_k(u,l),\;D_k(u,l)\subset DB_k(u,l)$,
of the billiards
which boundary $\p D_k(u,l)$ contains of course $A_k(u,l)$. The remaining part of $\p D_k(u,l)$ is built of the two
"limit" rays of $B_k(u,l)$ emerging from the ends of $A_k(u,l)$ and of $BA_k(u,l)$ if there is no caustic of the bundle
$B_k(u,l)$ inside the billiards or by the corresponding
caustic $K_k(u,l)=\{(f_k(K_k(s;u,l),s;u,l),\;g_k(K_k(s;u,l),s;u,l)):J_k(K_k(s;u,l),s;u,l)=0,\; {\bf r}_0(s)
\in A_k(u,l)\}$.

Similarly $\Psi_{A;k}^\sigma(d,s;u,l;\lambda),\;\sigma=\pm,$ is defined in the domain $D_k^A(u,l)$ corresponding to
the bundle $B_k^A(u,l)$.

An important property of the representation \mref{17} is its uniqueness, i.e. for two
different bundles defined on the segment $A_k(u,l)$
this representation provides us with two different pairs of $\Psi_k^\sigma(d,s;u,l;\lambda),\;\sigma=\pm$. This conclusion follows
from the fact that for $\lambda$ sufficiently large the BSWF's are determined only by exponentials and the latter are
different at the same points $(x,y)$ for different bundles.

\subsection{SWF's vanishing on the billiards boundary}

\hspace{15pt}Another obvious property of BSWF's $\Psi_k^\sigma(d,s;u,l;\lambda),\;\sigma=\pm$, is that they cannot vanish on $A_k(u,l)$
unless $\chi_k^\sigma(d,s;u,l;\lambda)$ vanish there
identically. Therefore a wave function $\Psi_k^{as;\sigma}(x,y;u,l;\lambda)$ vanishing on $A_k(u,l)$ should be represented in the
semiclassical limit by a linear combination of at least two BSWF's of the form \mref{17}. It is shown in App.A that the
proper linear combinations have to be the following:
\be
\Psi_k^{as;\sigma}(x,y;u,l;\lambda)=\Psi_{k}^{\sigma}(d_1,s_1;u,l;\lambda)+\Psi_{A;k}^{-\sigma}(d_2,s_2;u,l;\lambda)=\nn\\
J_{k}^{-\fr}(d_1,s_1;u,l)e^{\sigma i k\ll(d_1+\int_u^{s_1}\cos\alpha_{k}(s';u,l)ds'\r)}
\chi_{k}^{\sigma}(d_1,s_1;u,l;\lambda)+\nn\\
J_{A;k}^{-\fr}(d_2,s_2;u,l)e^{-\sigma i k\ll(d_2-\int_u^{s_2}\cos\alpha_{k}(s';u,l)ds'\r)}
\chi_{A;k}^{-\sigma}(d_2,s_2;u,l;\lambda)
\label{19}
\ee
with the following boundary conditions:
\be
\chi_{k}^{\sigma}(0,s;u,l;\lambda)+\chi_{A;k}^{-\sigma}(0,s;u,l;\lambda)=0
\;\;\;\;\;\;\; {\bf r}_0(s)\in A_k(u,l)
\label{19a}
\ee
while the point $(x,y)$ is the cross point of the respective trajectories belonging to different bundles, i.e.
\be
{\bf r}\equiv[x,y]={\bf r}_{k}(d_1,s_1;u,l)={\bf r}_0(s_1)+{\bf d}_{1}(s_1;u,l)=\nn\\
{\bf r}_{A;k}(d_2,s_2;u,l)={\bf r}_0(s_2)+{\bf d}_{2}(s_2;u,l)\nn\\
{\bf r}_{k}(d,s;u,l)\in B_k(u,l),\;\;\;\;\;{\bf r}_{A;k}(d,s;u,l)\in B_k^A(u,l)
\label{20}
\ee

The vanishing superposition \mref{19} if defined on the bundle $B_{k'}(u',l')$ suggests that
\linebreak $\Psi_{A;k'}^{-\sigma}(d,s;u',l';\lambda)$ should be related somehow to the BSWF
$\Psi_k^{\sigma}(d,h_k^{-1}(s;u,l);u,l);u,l;\lambda)$ defined on the bundle $B_{k}(u,l)$ which the previous one is
a reflection. In the next section this relation is established as a condition matching both the solutions.

\subsection{SWF's defined on a bundle skeleton and their continuity}

\hspace{15pt}According to our construction of the skeletons {\bf B}  and ${\bf B}^A$ there is a domain
$D_{\p B},\;B\supset D_{\p B}\supset\p B$,
of the billiards containing the billiards boundary $\p B$ in which each point $(x,y)$ with its some small vicinities is
mapped in the one-to-one way into each bundle of the pairs $B_k(u,l)$ and $B_k^A(u,l)$ containing this point.

Let ${\bf r}=(x,y)\in D_{\p B}$ be a fixed point of the billiards. Let $D(x,y)$ denote a set of all $D_k(u,l)$,\-
$B_k(u,l)\in{\bf B}$,
which contain this point and $D_A(x,y)$ is the respective set of $D_k^A(u,l),\; B_k^A(u,l)\in{\bf B}^A$. According to
their definition $D_k(u,l)\in D(x,y)$ if and only if $D_k^A(u,l)\in D_A(x,y)$.

SWF's $\Psi_{\bf B}^{as;\sigma}(x,y,\lambda)$ vanishing on the billiards boundary can now be defined on ${\bf B}$ and
${\bf B}^A$ in the domain $D_{\p B}$ as follows:
\be
\Psi_{\bf B}^{as;\sigma}(x,y,\lambda)=
\sum_{\ba{l}D_k(u,l)\in D(x,y)\\D_k^A(u,l)\in D_A(x,y)\ea}(\Psi_k^\sigma(d(u,l),s(u,l);u,l;\lambda)+\nn\\
\Psi_{A;k}^{-\sigma}(d(u,l),s(u,l);u,l;\lambda))
\label{23}
\ee
where BSWF's $\Psi_k^\sigma(d(u,l),s(u,l);u,l;\lambda)$ and $\Psi_{A;k}^{-\sigma}(d(u,l),s(u,l);u,l;\lambda)$ are
defined in $D_{\p B}$ and  satisfy the condition \mref{19a} on $A_k(u,l)$.

The solutions $\Psi_{\bf B}^{as;+}(x,y,\lambda)$ and $\Psi_{\bf B}^{as;-}(x,y,\lambda)$ coincide if and only if
${\bf B}={\bf B}^A$.

The SWF's \mref{23} which satisfy the condition of vanishing on the billiards boundary are the most general
ones for the skeletons {\bf B} and ${\bf B}^A$ which can be defined in the domain $D_{\p B}$. However the next step in
solving the basic problem of energy
quantization in the semiclassical limit is to make these solutions continuous in $D_{\p B}$ since
this property is not ensured automatically by \mref{23}. $\Psi_{\bf B}^{as;\sigma}(x,y,\lambda)$ are certainly continuous and
unique inside each bundle contained in $D_{\p B}$. However if the skeleton {\bf B} is singular then their bundles have
their boundaries also inside the billiards area on which $\Psi_{\bf B}^{as;\sigma}(x,y,\lambda)$ or their derivatives
can appear to be discontinuous if a point $(x,y)$
crosses these boundaries. They can be also non unique if a point $(x,y)$ moves along some closed loops such as the one which is
homotopic with the billiards boundary, i.e. it can happen that
$\Psi_{\bf B}^{as;\sigma}(x,y,\lambda)$ do not come back to their initial values being continued along such a loop.

Considering the continuity property of $\Psi_{\bf B}^{as;\sigma}(x,y,\lambda)$ the following circumstances can
accompany in general such bundle boundary crossings:
\begin{enumerate}
\item two BSWF's $\Psi_k^\pm(d,s;u,l;\lambda)$ and $\Psi_{k'}^\pm(d',s';u',l';\lambda)$ which enter the sum \mref{23}
are defined on the bundles $B_k(u,l)$ and $B_{k'}(u',l')$ which can be composed into a compound bundle; and
\item there is no such BSWF's and the respective neighboring bundles.
\end{enumerate}

In the first of the above cases the corresponding $\Psi_k^\pm(d,s;u,l;\lambda),\;(x_k(d,s),y_k(d,s))\in DB_k(u,l)$, and
$\Psi_{k'}^\pm(d',s';u',l';\lambda),\;(x_{k'}(d',s'),y_{k'}(d',s'))\in DB_{k'}(u',l')$, defined by \mref{17A} have to
be identified on the common boundary of the bundles $B_k(u,l)$ and $B_{k'}(u',l')$ together with their first derivatives,
i.e.
\be
\Psi_k^\pm(d,s;u,l;\lambda)=\Psi_{k'}^\pm(d',s';u',l';\lambda)\nn\\
\frac{\p}{\p x}\Psi_k^\pm(d,s;u,l;\lambda)=\frac{\p}{\p x}\Psi_{k'}^\pm(d',s';u',l';\lambda)\nn\\
\frac{\p}{\p y}\Psi_k^\pm(d,l;u,l;\lambda)=\frac{\p}{\p y}\Psi_{k'}^\pm(d',s';u',l';\lambda)\nn\\
(x_k(d,l),y_k(d,l))\equiv (x_{k'}(d',s'),y_{k'}(d',s'))\in \p DB_k(u,l)\cap\p DB_{k'}(u',l')\neq\oslash
\label{23a}
\ee

Note that such identifications as the last ones mean that the BSWF's are defined now on the reduced skeleton ${\bf B}^R$
rather then on the original ones.

In the second case however the corresponding BSWF's $\Psi_k^\pm(d,s;u,l;\lambda),\;(x(d,s),y(d,s))\in DB_k(u,l)$
or their normal derivatives have to vanish on such a boundary of $DB_k(u,l)$, i.e.
\be
\Psi_k^\pm(d,s;u,l;\lambda)=0\nn\\
(x_k(d,s),y_k(d,s))\in\p DB_k(u,l)
\label{23b}
\ee
or
\be
\frac{\p}{\p n}\Psi_k^\pm(d,s;u,l;\lambda)=0\nn\\
(x(d,s),y(d,s))\in\p DB_k(u,l)
\label{23ba}
\ee
that is in such cases $\Psi_k^\pm(d,s;u,l;\lambda)$ defined in the bundle $B_k(u,l)$ should satisfy on its boundary
Dirichlet's or Neumann's conditions.

The last two conditions though necessary seem to look as a little bit arbitrary. However we should remember that our
calculations are performed in the semiclassical regime, i.e. in the classically allowed regions (bundles) outside which
the semiclassical wave functions cannot exist. Physically this means obviously that outside each bundle a
corresponding piece of the exact wave function represented on the bundle by its respective semiclassical approximation
has to vanish exponentially (for $\lambda$ sufficiently large) when moving away from the bundle. Semiclassically it
just means that BSWF's defined inside the bundles have to vanish identically outside of them. This condition can course
however that the first derivatives of the global SWF's \mref{23} can be discontinuous on such bundles boundaries. Just
this last property differs essentially the semiclassical solutions \mref{23} from the exact ones. If it happens we will
call such a GSWF {\bf singular} in  contrast to the {\bf regular} one which is continuous in the whole billiards
together with its first derivatives. From this discussion it follows also that for the latter possibility to happen
it is necessary for the skeleton on which $\Psi_{\bf B}^{as;\sigma}(x,y,\lambda)$ is defined to be global.

\subsection{First quantization condition for SWF's}

\hspace{15pt}If $\Psi_{\bf B}^{as;\sigma}(x,y,\lambda)$ are made continuous inside $D_{\p B}$ then the uniqueness
condition for
them on each loop lying in $D_{\p B}$ particularly on the ones homotopic with the billiards boundary $\p B$ if such exists
leads us to the first quantization condition which has to be satisfied by these two SWF's.

\subsection{Continuing the SWF's $\Psi_{\bf B}^{as;\sigma}(x,y,\lambda)$ over the whole skeletons
{\bf B} and ${\bf B}^A$ - the global SWF's}

\hspace{15pt}By the formula \mref{23} $\Psi_{\bf B}^{as;\sigma}(x,y,\lambda)$ are defined in every bundle of the skeletons
{\bf B} and ${\bf B}^A$ close to the billiards boundary. However by the way of construction of both the skeletons
if a point $(x,y)$ of the domain $D_{\p B}$ is achieved by a ray of some bundle of {\bf B} running all the time by the
domain $D_{\p B}$ then it is also achieved by the same ray inverted in time and being a member of a bundle of ${\bf B}^A$,
i.e. running in the opposite direction. However the second ray to achieve the considered point $(x,y)\in D_{\p B}$ has
in general first to leave the domain $D_{\p B}$ crossing its boundary in several points in order to come back to it.

The same note is valid for rays contained in the skeleton ${\bf B}^A$.

Therefore $\Psi_{\bf B}^{as;\sigma}(x,y,\lambda)$ defined by the formula \mref{23} can be continued from a point $(x,y)$ of the domain $D_{\p B}$
into another such point of $D_{\p B}$ along rays contained in the skeletons {\bf B} or ${\bf B}^A$. If $D_{\p B}$
cannot be equal $B$, then such a continuation meet as necessary caustic points which have to be avoided somehow.
If we do that however we will achieve again points of the domain
$D_{\p B}$ and naturally the
continued solutions and the solutions defined by \mref{23} have to coincide. This coincidence formulate the second
quantization condition which both the SWF's $\Psi_{\bf B}^{as;\sigma}(x,y,\lambda)$ have to satisfy. Such a coincidence
is achieved by identifying each term of the sum \mref{23} with the corresponding term of the continued
$\Psi_{\bf B}^{as;\sigma}(x,y,\lambda)$. Anticipating the results of App.B of \cite{3} the corresponding identification
should be done as follows.

\begin{enumerate}
\item Let $B_k(u,l)$ be a reflection of the bundles $B_{k_1}(u_1,l_1),B_{k_2}(u_2,l_2),\ldots,B_{k_n}(u_n,l_n)$ satisfying \mref{16a}.
Let $\Psi_{k_1}^{\sigma;cont}(d,s;u_1,l_1;\lambda),\Psi_{k_2}^{\sigma;cont}(d,s;u_2,l_2;\lambda),\ldots,
\Psi_{k_n}^{\sigma;cont}(d,s;u_n,l_n;\lambda)$ denote the SWF's defined in $D_{\p B}$ and continued on the respective
bundles $B_{k_1}(u_1,l_1),\linebreak B_{k_2}(u_2,l_2),\;\ldots,B_{k_n}(u_n,l_n)$ again to $D_{\p B}$. Let further
$\Psi_{A;k}^{-\sigma}(d,s;u,l;\lambda)$ be defined in $D_{\p B}$ on the bundle $B_k^A(u,l)$ while
$\Psi_k^{\sigma}(d,s;u,l;\lambda)$ in $D_{\p B}$ on the bundle $B_k(u,l)$ being both related by the boundary condition
\mref{19a}.

Then we make the following identification of BSWF's:
\be
\Psi_{A;k}^{-\sigma}(d,h(s;u_j,l_j);u,l;\lambda)=
\Psi_{k_j}^{\sigma;cont}(D(s;u_j,l_j)-d,s;u_j,l_j;\lambda)\nn\\
{\bf r}_0(h(s;u_j,l_j))\in A_k(u,l)\cap BL(u_j,l_j)\nn\\
{\bf r}_0(s)\in L(u_j,l_j)\nn\\
{\bf r}_0(h(s;u_j,l_j))={\bf r}_0(s)+{\bf D}(s;u_j,l_j),\;\;\;\;\;j=1,...n
\label{21}
\ee
where $D(s;u,l)=|{\bf D}(s;u,l)|$ denotes the distance between the points ${\bf r}_0(h_k(s;u,l))$ and
${\bf r}_0(s)$ of the boundary $\p B$.

Similarly
\item Let $B_k^A(u,l)$ be a reflection of the bundles $B_{j_1}^A(u_1',l_1'),B_{j_2}^A(u_2',l_2'),\ldots,B_{j_m}^A(u_m',l_m')$ satisfying \mref{16a}.
Let $\Psi_{A;j_1}^{\sigma;cont}(d,s;u_1',l_1';\lambda),\Psi_{A;j_2}^{\sigma;cont}(d,s;u_2',l_2';\lambda),\ldots,
\Psi_{A;k_n}^{\sigma;cont}(d,s;u_m',l_m';\lambda)$ denote the SWF's defined in $D_{\p B}$ and continued on the respective
bundles $B_{j_1}^A(u_1',l_1'),\linebreak B_{j_2}^A(u_2',l_2'),\ldots,B_{j_m}^A(u_m',l_m')$ again to $D_{\p B}$. Let further
$\Psi_k^{-\sigma}(d,s;u,l;\lambda)$ be defined in $D_{\p B}$ on the bundle $B_k(u,l)$ while
$\Psi_{A;k}^{\sigma}(d,s;u,l;\lambda)$ in $D_{\p B}$ on the bundle $B_k^A(u,l)$ being both related by the boundary condition
\mref{19a}.

Then we make the following identification:
\be
\Psi_k^{-\sigma}(d,h(s;u_i,l_i);u,l;\lambda)=
\Psi_{A;j_i}^{\sigma;cont}(D(s;u_i,l_i)-d,s;u_i,l_i;\lambda)\nn\\
{\bf r}_0(h(s;u_i,l_i))\in A_k(u,l)\cap BL(u_i,l_i)\nn\\
{\bf r}_0(s)\in L(u_i,l_i)\nn\\
{\bf r}_0(h(s;u_i,l_i))={\bf r}_0(s)+{\bf D}(s;u_i,l_i),\;\;\;\;\;i=1,...m
\label{22}
\ee

\item Meeting the caustic points the BSWF's $\Psi_k^\sigma(d,s;u,l;\lambda)$ and $\Psi_{A;k}^\sigma(d,s;u,l;\lambda)$
avoid them by fixing $s$ and moving on the complex $d$-plane from above the points for $\sigma=+$ and from below them
for $\sigma=-$.
\end{enumerate}

The conditions \mref{21} - \mref{22} allow us to define $\Psi_{\bf B}^{as;\sigma}(x,y,\lambda)$ as given by \mref{23}
in every point of the domain $D_{\bf B}$, i.e. in the domain classically allowed when moving on the skeleton {\bf B} and
to rewrite the sum in \mref{23} representing $\Psi_{\bf B}^{as;\sigma}(x,y,\lambda)$ globally by the terms of
$\Psi_k^\sigma(d(u,l),s(u,l);u,l;\lambda)$ or by the terms of $\Psi_{A;k}^{-\sigma}(d(u,l),s(u,l);u,l;\lambda)$. Namely,
the global SWF's (GSWF) $\Psi_{\bf B}^{as;\sigma}(x,y,\lambda)$ are given by:
\be
\Psi_{\bf B}^{as;\sigma}(x,y,\lambda)=
\sum_{DB_k(u,l)\in {\tilde D}(x,y)}\Psi_k^\sigma(d(u,l),s(u,l);u,l;\lambda)=\nn\\
\sum_{DB_k^A(u,l)\in {\tilde D}_A(x,y)}\Psi_{A;k}^{-\sigma}(d(u,l),s(u,l);u,l;\lambda)
\label{21a}
\ee
where ${\tilde D}(x,y)$ and ${\tilde D}_A(x,y)$ denote now the respective sets of $DB_k(u,l)$ and $DB_k^A(u,l)$
containing the point $(x,y)$.

The sums in \mref{21a} contain all BSWF's $\Psi_k^\sigma(d,s;u,l;\lambda)$ and
$\Psi_{A;k}^{-\sigma}(d,s;u,l;\lambda)$ which can be continued to this point by the corresponding
domains $DB_k(u,l)$ and $DB_k^A(u,l)$.

Let us note that the forms \mref{21a} of the GSWF's allow us in fact to define them on the reduced form
${\bf B}^R$ of the skeleton {\bf B} rather then on the skeleton itself. This possibility permits to reduce substantially
number of terms in sums \mref{21a}.

Rewritten in terms of the $\chi$-coefficients Eq.\mref{21} gives:
\be
\chi_{A;k}^{-\sigma}(d,h(s;u_j,l_j);u,l;\lambda)=\nn\\
\eta_\sigma e^{\sigma i\lambda p\delta_k(u_j,l_j)}
\ll|\frac{\p h(s;u_j,l_j)}{\p s}\r|^{-\fr}\chi_{k_j}^{\sigma;cont}(D(s;u_j,l_j)-d,s;u_j,l_j;\lambda)\nn\\
\delta_k(u_j,l_j)=D(s;u_j,l_j)+\int_{u_j}^s\cos\alpha_{k_j}(s';u_j,l_j)ds'-
\int_u^{h(s;u_j,l_j)}\cos\alpha_k(s';u,l)ds'\nn\\
{\bf r}_0(h(s;u_j,l_j))\in A_k(u,l)\cap BL(u_j,l_j)\nn\\
{\bf r}_0(s)\in L(u_j,l_j)\nn\\
{\bf r}_0(h(s;u_j,l_j))={\bf r}_0(s)+{\bf D}(s;u_j,l_j),\;\;\;\;\;j=1,...n
\label{22a}
\ee

Note that $\delta_k(u_j,l_j)$ in the above formula is $s$-independent (see App.B). Due to that and due to the
properties \mref{13a} and \mref{13b} the rhs of \mref{22a} satisfies \mref{11} as it should.

Putting $d=0$ in \mref{22a} and taking into account \mref{19a} we get:
\be
\chi_k^{\sigma}(0,h(s;u_j,l_j);u,l;\lambda)=-\chi_{A;k}^{-\sigma}(0,h(s;u_j,l_j);u,l;\lambda)=\nn\\
-\eta_\sigma e^{\sigma i\lambda p\delta_k(u_j,l_j)}\ll|\frac{\p h(s;u_j,l_j)}{\p s}\r|^{-\fr}
\chi_{k_j}^{\sigma;cont}(D(s;u_j,l_j),s;u_j,l_j;\lambda)=\nn\\
-\eta_\sigma e^{\sigma i\lambda p\delta_k(u_j,l_j)}\chi_{k_j}^{\sigma,cont}(0,h(s;u_j,l_j);u,l;\lambda)\nn\\
{\bf r}_0(h(s;u_j,l_j))\in A_k(u,l)\cap BL(u_j,l_j)\nn\\
{\bf r}_0(s)\in L(u_j,l_j)\nn\\
{\bf r}_0(h(s;u_j,l_j))={\bf r}_0(s)+{\bf D}(s;u_j,l_j),\;\;\;\;\;j=1,...n
\label{22b}
\ee

The BSWF's $\Psi^\sigma(d,s;u,l;\lambda)$ defined on bundles of ${\bf B}$ and $\Psi_A^\sigma(d,s;u,l;\lambda)$ defined on
respective bundles of ${\bf B}^A$ are related with each other by the boundary conditions \mref{19a} and by matching conditions
\mref{22}-\mref{22b}.

It is clear that the conditions \mref{22b} have to determine also the $\chi$-factors $\chi_k^\sigma(s;u,l;\lambda)$
for all the bundles $B_k(u,l)$ which are the "initial" conditions for both $\chi_k^\sigma(d,s;u,l;\lambda)$ and
$\chi_{A;k}^\sigma(d,s;u,l;\lambda)$ in the recurrent formula \mref{12z}, i.e.
$\chi_k^\sigma(s;u,l;\lambda)\equiv\chi_k^\sigma(0,s;u,l;\lambda)\equiv$\linebreak
$-\chi_{A;k}^\sigma(0,s;u,l;\lambda)$.  Nevertheless these conditions
cannot be given arbitrarily. Just opposite all $\chi_k(s;u,l;\lambda)$ have to satisfy \mref{22b} in a selfconsistent
way.

The formulae \mref{22a} and \mref{22b} define the conditions which the SWF's
$\chi_k^{\sigma}(d,h(s;u_j,l_j);u,l;\lambda)$ should satisfy when bouncing from the billiards boundary. Nevertheless
this condition can be specified additionally with respect to its factors. Namely, taking their large $\lambda$-limit
we get:
\be
\chi_{k,0}^{\sigma}(h(s;u_j,l_j);u,l)=-\eta_\sigma \ll|\frac{\p h(s;u_j,l_j)}{\p s}\r|^{-\fr}
e^{\sigma i\lambda p\delta_k(u_j,l_j)}\chi_{k_j,0}^{\sigma}(s;u_j,l_j)\nn\\
\chi_{k,r+1}^{\sigma}(h(s;u_j,l_j);u,l)=-\eta_\sigma \ll|\frac{\p h(s;u_j,l_j)}{\p s}\r|^{-\fr}
e^{\sigma i\lambda p\delta_k(u_j,l_j)}\times\nn\\
\ll(\chi_{k_j,r+1}^{\sigma}(s;u_j,l_j)+\frac{\sigma i}{2p}\int_0^{D(s;u_j,l_j)}\ll(\frac{{}^{}}{{}_{}}J^{\fr}\triangle\ll(J^{-\fr}\chi_{k_j,r}^\sigma(a,s;u_j,l_j)\r)\r.\r.+\nn\\
\ll.\ll.2\sum_{l=0}^rE_{r-l+1}\chi_{k_j,l}^\sigma(a,s;u_j,l_j)\r)da\r)\nn\\
r=0,1,2,...,
\label{22c}
\ee

The above equations should be satisfied on each bundle $B_k(u,l)$ of the skeleton ${\bf B}$.

The first of the equations \mref{22c} should determine the classical quantities, namely the skeleton {\bf B} and the
"classical" energy $E_0=\fr p^2$ and by them define the JWKB approximation of the SWF's. Namely:
\be
\Psi_{\bf B}^{JWKB;\sigma}(x,y,\lambda)=\sum_{DB(u,l)\in D(x,y)}\Psi^{JWKB;\sigma}(d(u,l),s(u,l);u,l)=\nn\\
\sum_{DB(u,l)\in D(x,y)}J^{-\fr}(d(u,l),s(u,l))
e^{\sigma i\lambda p\ll(d(u,l)+\int_{u}^{s(u,l)}\cos\alpha(s';u,l)ds'\r)}\chi_0^\sigma(d(u,l),s(u,l);u,l)
\label{22e}
\ee

The remaining equations determine quantum corrections to the "classical" ones involved in \mref{22e}.

However it is easy to note that for the selfconsistency of the equations \mref{22c} it is necessary for the exponent
$e^{\sigma i\lambda p\delta_k(u_j,l_j)}$ to be independent of $\lambda$, i.e. we have to have on each bundle $B_k(u,l)$
of {\bf B}:
\be
\lambda p\delta_k(u,l)=\phi_k(u,l)\nn\\
B_k(u,l)\subset {\bf B}
\label{22d}
\ee
where $\delta_k(u,l)$ is given by \mref{22a} and $\phi_k(u,l)$ is a $\lambda$-independent constant.

The equations \mref{22d} have to define the energy $E_0=\fr p^2$.

Taking into account the last conclusions we get the following final set of the recurrent quantization conditions:
\be
\lambda p\delta_k(u,l)=\phi_k(u,l)\nn\\
\chi_{k,0}^{\sigma}(h(s;u_j,l_j);u,l)=-\eta_\sigma e^{\sigma i\phi_{k_j}(u_j,l_j)}\ll|\frac{\p h(s;u_j,l_j)}{\p s}\r|^{-\fr}\chi_{k_j,0}^{\sigma}(s;u_j,l_j)\nn\\
\chi_{k,r+1}^{\sigma}(h(s;u_j,l_j);u,l)=
-\eta_\sigma  e^{\sigma i\phi_{k_j}(u_j,l_j)}\ll|\frac{\p h(s;u_j,l_j)}{\p s}\r|^{-\fr}
\ll(\frac{{}^{}}{{}_{}}\chi_{k_j,r+1}^{\sigma}(s;u_j,l_j)+\r.\nn\\
\ll.\frac{\sigma i}{2p}\int_0^{D(s;u_j,l_j)}\ll(\frac{{}^{}}{{}_{}}J^{\fr}\triangle\ll(J^{-\fr}\chi_{k_j,r}^\sigma(a,s;u_j,l_j)\r)+
2\sum_{l=0}^rE_{r-l+1}\chi_{k_j,l}^\sigma(a,s;u_j,l_j)\r)da\r)\nn\\
r=0,1,2,...,
\label{22f}
\ee
together with:
\be
\chi_{k,0}^{\sigma}(d,s;u,l)\equiv \chi_{k,0}^{\sigma}(s;u,l)\nn\\
\chi_{k,r+1}^{\sigma}(d,s;u,l)=\chi_{k,r+1}^{\sigma}(s;u,l)+\nn\\
\frac{\sigma i}{2p}\int_0^d
\ll(J^{\fr}\triangle\ll(J^{-\fr}\chi_{k,r}^\sigma(a,s;u,l)\r)+
2\sum_{m=0}^rE_{r-l+1}\chi_{k,m}^\sigma(a,s;u,l)\r)da\nn\\
r=0,1,2,...,
\label{22g}
\ee

Let us note finally that if ${\bf B}\neq{\bf B}^A$ then energy levels corresponding to the skeleton ${\bf B}$ have
to be degenerate. This conclusion follows easily from the form of the quantization conditions \mref{22f}-\mref{22g} and
\mref{23a}-\mref{23b} showing that the complex conjugations of $\Psi_{\bf B}^{as;\sigma}(x,y,\lambda)$ satisfy also
these conditions with the same semiclassical energy $E$. The two corresponding solutions are of course
$\Psi_{\bf B}^{as;\pm}(x,y,\lambda)$.

\subsection{Finite and infinite bundle structures of skeletons. The last quantization condition}

\hspace{15pt}The constructions of skeletons and SWF's in billiards performed in sec.3-4 describe completely
the energy quantization problem in the semiclassical approximation.

For a given billiards however there can be skeletons with a finite number of bundles as well as with an infinite
one. The semiclassical quantization procedure described in the previous sections seems to be easily applied to
the finite bundle number skeletons. Namely in such a case following a trajectory starting from a bundle $B_k(u,l)$ we have
to approach the same bundle after a finite number of bounces. The corresponding semiclassical wave function propagated
by the skeleton has therefore to come back to its initial form achieving again the initial bundle. This condition
closes essentially the process of quantization formulated in the previous sections. The respective conditions are of course the
following:
\be
\exp\ll(\sum_{B_k(u,l)\in {\bf B}}\delta_k(u,l)\r)\prod_{k=1}^n(-\eta_{\sigma_k})=1\nn\\
\chi_k^{\sigma,cont}(D(s,s'),s;u,l;\lambda)=\chi_k^\sigma(s';u,l;\lambda),\;\;\;\;\;\;s,s'\in B_k(u,l)
\label{23ca}
\ee
where $n$ is a number of bounces and $D(s,s')$ is the global distance passed by the billiards ball along the investigated
trajectory.

Skeletons with a finite number of bundles are typical for the billiards with the integrable or pseudointegrable motions.
Nevertheless they can be found also as particular cases of motions in chaotic billiards as well.

The cases of skeletons with an infinite number of bundles are clearly much more difficult for investigations. Such skeletons
should be typical for chaotic billiards.

According to its definition a bundles $B_k(u,l)$ can bifurcate after
the reflection by the billiards boundary into many different subbundles, i.e. parts of other bundles having
their beginnings also partly on the arc $BA_k(u,l)$. In fact a general behaviour of a skeleton in such chaotic cases
should not
differ essentially by its chaotic complexity from a chaotic trajectory reminding however rather a gigantic road-knot with
infinitely many viaducts spanning the billiards boundary on which the billiards ball moves. It is obvious that if they
exist their identification seems to be not an easy task.

Nevertheless the rule \mref{23c} can appear to be useful also even in such cases. This is because a ray beginning with
a bundle $B_k(u,l)$ can come back to it even arbitrarily close to its initial starting point on $A_k(u,l)$ (according to
the Poincare theorem) not repeating its way. But this is enough for writing the "last quantization condition" \mref{23c}
where the sum goes now over all bundles of the skeleton passed by the ray.

In the next two sections we shall focus on the finite number cases of bundles in skeletons, i.e. applying this
procedure to the simplest well known cases of the polygon billiards not avoiding however
billiards with chaotic motions such as the Bunimovich one.

\section{The rational polygon billiards}

\hskip+2em A two dimensional rational polygon billiards are distinguished by their pseudointegrability \cite{8}. As we
have discussed it in sec.3 a phase space
corresponding to a motion in such a billiards on a given skeleton consists of a finite number of pieces parallel to the
billiards plane and
orthogonal to the two momentum axes and corresponding each to the compound bundles the reduced skeleton. In the case
when the corresponding skeleton is regular then by the gluing procedure described in sec.3.2 one
can get \cite{8,6} a two dimensional compact closed surface with a genus $g$ given by:
\be
g=1+\frac{N}{4}\sum_{k=1}^n\frac{p_k-1}{q_k}
\label{23c}
\ee
where $N$ is the number of the compound bundles, $n$ is the number of the polygon vertices, and
$\pi\frac{p_k}{q_k}$ with integers $p_k,\;q_k$ relatively prime is the angle enclosed $k$-th vertex, $k=1,...,n$.

In other cases of the skeletons developed in the rational polygons one gets surfaces which do not provide us with closed
surfaces in the phase space, i.e. such skeletons are singular. In particular such singular skeletons are developed by
periodic trajectories.

Note that it is the polygon skeleton property that if it contains at least one periodic trajectory then
all trajectories of such a skeleton are also periodic. In such a polygon periodic skeleton there are always
two (and no more) periodic trajectories each of which starts from some vertex of the polygon and runs to another one.
These two periodic trajectories have been called by Bogomolny and Schmit \cite{1} as {\bf singular diagonals} (SD's)
while the skeleton itself as the periodic orbit channel (POC). Therefore each periodic skeleton is defined by two SD's.

A convenient way of representing motions in a polygon billiards can be obtain by unfolding the polygon by its
repeating reflections in its sides on which the trajectory reflections are performed. A frequently complicated pattern
of the real trajectories takes then a simple form of parallel straight lines on such unfolded polygons.

While a triangle is the simplest polygon its billiards properties are in general not as such. A motion in rational
triangles can be integrable if $g=1$ for them so that for the triangle angles $\pi\frac{p_i}{q_i},\;i=1,2,3,$ as it
follows from \mref{23c} we have to have:
\be
p_i=1,\;\;\;\;\;\;\;i=1,2,3\nn\\
\frac{1}{q_1}+\frac{1}{q_2}+\frac{1}{q_3}=1
\label{23d}
\ee

Several obvious solutions to \mref{23d} give the following triangle angles $\ll(\frac{\pi}{2},\frac{\pi}{3},\frac{\pi}{6}\r)$,
$\ll(\frac{\pi}{2},\frac{\pi}{4},\frac{\pi}{4}\r)$ and $\ll(\frac{\pi}{3},\frac{\pi}{3},\frac{\pi}{3}\r)$ for the integrable cases.

In fact it is rather a rectangular billiards and its variations which we call
broken rectangle billiards which seems to demonstrate sometimes in a spectacular way most advantages of the skeleton approach
developed in sec.3. Therefore we will firstly consider the cases just mentioned. The cases of the equilateral triangle and the
pentagon billiards will be considered next.

\subsection{The rectangular billiards}

\hskip+2em Consider therefore the rectangular billiards shown in Fig.2. This billiards is the canonical example of the
energy quantization problem because of its easiness to be solved by the variable separation method. According to
Fig.2 the well known solution to the problem is given by the following two equations:
\be
\lambda p_xa=m\pi\nn\\
\lambda p_yb=n\pi\nn\\
m,n=1,2,3,...
\label{A0}
\ee
giving the energy:
\be
E=\fr p_x^2+\fr p_y^2=\frac{\pi^2}{2\lambda^2}\ll(\frac{m^2}{a^2}+\frac{n^2}{b^2}\r)
\label{A0f}
\ee
and being the result of the following form of the (non-normalized) energy eigenfunctions:
\be
\Psi(x,y)=4\sin(\lambda p_xx)\sin(\lambda p_yy)=\nn\\
e^{\lambda p_xx-\lambda p_yy}+e^{-\lambda p_xx+\lambda p_yy}-e^{\lambda p_xx+\lambda p_yy}-
e^{-\lambda p_xx-\lambda p_yy}
\label{A0a}
\ee
which have to vanish on the billiards boundary.

Of course one can always put $p_x=p\cos\alpha,p_y=p\sin\alpha$ where $\alpha,\;0<\alpha<\fr\pi$, is the
angle by which the momentum $p$ is inclined to the $x$-axis when the billiards ball reflects from the side $A_1$.
Therefore the classical trajectory angles of the billiards ball are quantized according to the formula:
\be
\tan\alpha=\frac{n}{m}\frac{a}{b}\nn\\
m,n=1,2,3,...
\label{A0e}
\ee

Let us note that the cases $\alpha=0,\fr\pi$ are excluded by the solutions \mref{A0a}.

Let us note further that the set $\Sigma$ of all pairs $(m,n),\;m,n=1,2,3,...,$ defining the eigenfunctions
$\Psi_{m,n}(x,y)$ can be divided into disjoint subsets $\Sigma_{m_0,n_0}$ each of which contains a pair $(m_0,n_0)$
where $m_0$ and $n_0$ are relatively prime and all its multiples $(km_0,kn_0),\;k=1,2,3,...,$ and nothing more so that
$\Sigma=\bigcup_{m_0,n_0}\Sigma_{m_0,n_0}$. It is clear that all the points of $\Sigma_{m_0,n_0}$ lie on the straight
line $y=\tan\alpha_0x$ with $\tan\alpha_0=\frac{n_0}{m_0}\frac{a}{b}$, i.e. all the states
$\Psi_{km_0,kn_0}(x,y),\;k=1,2,3,...,$ are related in the rectangular billiards with a family of classical trajectories
which are inclined to the $x$-axis by the angle $\alpha_0$.

\begin{figure}
\begin{center}
\psfig{figure=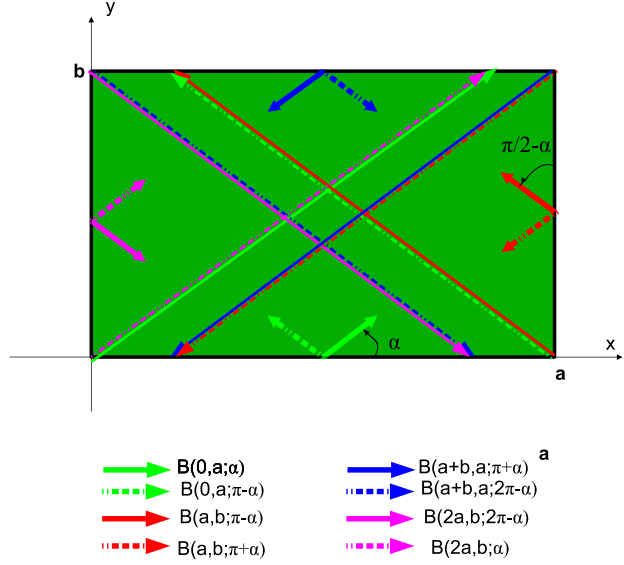,width=9.5cm}
\caption{The eight ray bundles of the generic regular skeletons ${\bf B}$ and ${\bf B}^T(\equiv{\bf B})$ in the rectangular
billiards and the one ($B_8\cup B_1$) of the four compound bundles. The latter form a torus in the phase space.}
\end{center}
\end{figure}

\subsection{The rectangular billiards skeletons built by non-periodic trajectories}

\hskip+2em To perform semiclassical calculations corresponding to "generic" skeletons let us consider a skeleton
shown in Fig.2 containing, by assumption, only nonperiodic
trajectories. According to the description of the previous section there are four
"smooth arcs" in the rectangular billiards, i.e. the four sides of the rectangle $A_1,...,A_4$. Since the absolute
values of the momentum components $p_x,p_y$ are the integrals of the classical motion inside the billiards respecting
elastic law of bouncing then all bundles which should be taken into account are defined by a single angle
$\alpha,\;0<\alpha<\fr\pi$, which are made by the rays of the bundle $B_1=B_1(0,a;\alpha)$ with the $x$-axis.

Choosing the case of the angle $\alpha$ shown in Fig.2 the remaining seven bundles of the skeleton {\bf B} shown in
this figure are:
$B_2=B_1(0,a;\pi-\alpha),\;B_3=B_2(a,b;\pi-\alpha),\;B_4=B_2(a,b;\pi+\alpha),\;B_5=B_3(a+b,a;\pi+\alpha),\;B_6=B_3(a+b,a;2\pi-\alpha),\;
B_7=B_4(2a+b,b;2\pi-\alpha),\;B_8=B_4(2a+b,b;\alpha)$, i.e. the parameter $s$ introduced in sec.3 is counted anticlockwise
starting from the
point $(0,0)$ of Fig.2 (and having negative value if measured clockwise). The bundles $B_{2k-1},\;B_{2k},\;k=1,...,4,$
are defined on the respective sides $A_k,\;k=1,...,4,$ of
the billiards, i.e. on $A_1=A_1(0,a),\;A_2=A_2(a,b),\;A_3=A_3(a+b,a),\;A_4=A_4(2a+b,b)$.

The skeleton ${\bf B}^T$ coincides exactly with {\bf B} in the case of generic skeletons in the rectangular
billiards, i.e. the corresponding energy levels cannot be degenerate.

Let us note that a number of bundles in the skeletons is obviously independent of a choice of $\alpha$, i.e. it is
always equal to eight if $0<\alpha<\fr\pi$.

The corresponding Jacobean factors of $\Psi_q^\pm(d,s,\lambda),\;q=1,...,8,$ are $J^{-\fr}(d,s)\equiv
(-\sin\alpha_k)^{-\fr}$, $\alpha_k=\alpha,\;
s\in A_k,\;k=1,3$ and $\alpha_k=\fr\pi-\alpha,\; s\in A_k,\;k=2,4$, i.e. the Jacobeans are
constant but discontinues. Therefore they will be included into the $\chi$-factors contained in the SWF's.

We can now make use of the fact that from the sixteen BSWF's $\Psi_q^\pm(d,s,\lambda),\;q=1,...,8,$ we can first select
only eight of them with the positive signature since the negative signature solutions have to coincide with the
respective positive signature ones. Next since each pair $\Psi_{2q}^+(d,s,\lambda),\;\Psi_{2q+1}^+(d,s,\lambda)\;q=1,...,4,$
of these solutions has to coincide on the common boundary of the respective bundles  $B_{2q}$ and $B_{2q+1},\;q=1,...,4,$
then we can define the solutions on the respective compound bundles to get in this way only four BSWF's,
namely:
\be
\ba{ll}
{\tilde\Psi}_1^+(d,s,\lambda)&\equiv\ll\{\ba{ll} \Psi_8^+(d,s,\lambda)&-b<s<0\\
                                                 \Psi_1^+(d,s,\lambda)&0<s<a
                                                 \ea\r.\\
{\tilde\Psi}_2^+(d,s,\lambda)&\equiv\ll\{\ba{ll} \Psi_2^+(d,s,\lambda)&0<s<a\\
                                                 \Psi_3^+(d,s,\lambda)&a<s<a+b
                                                 \ea\r.\\
{\tilde\Psi}_3^+(d,s,\lambda)&\equiv\ll\{\ba{ll} \Psi_4^+(d,s,\lambda)&a<s<a+b\\
                                                 \Psi_5^+(d,s,\lambda)&-a-b<s<-b
                                                 \ea\r.\\
{\tilde\Psi}_4^+(d,s,\lambda)&\equiv\ll\{\ba{ll} \Psi_6^+(d,s,\lambda)&-a-b<s<-b\\
                                                 \Psi_7^+(d,s,\lambda)&-b<s<0
                                                 \ea\r.\ea
\label{A0b}
\ee
and also the respective four compound bundles:
\be
{\tilde B}_1\equiv B_{8,1}=B_8\cup B_1\nn\\
{\tilde B}_2\equiv B_{2,3}=B_2\cup B_3\nn\\
{\tilde B}_3\equiv B_{4,5}=B_4\cup B_5\nn\\
{\tilde B}_4\equiv B_{6,7}=B_6\cup B_7\nn\\
\label{A0c}
\ee
on which the four solutions \mref{A0b} are defined.

Therefore the reduced skeleton ${\bf B}^R$ contains four compound bundles \mref{A0c}. Note that each compound bundle
${\tilde B}_k,\;k=1,...,4$, is regular so that the "generic" rectangular billiards skeleton {\bf B} is also regular.

\begin{figure}
\begin{center}
\psfig{figure=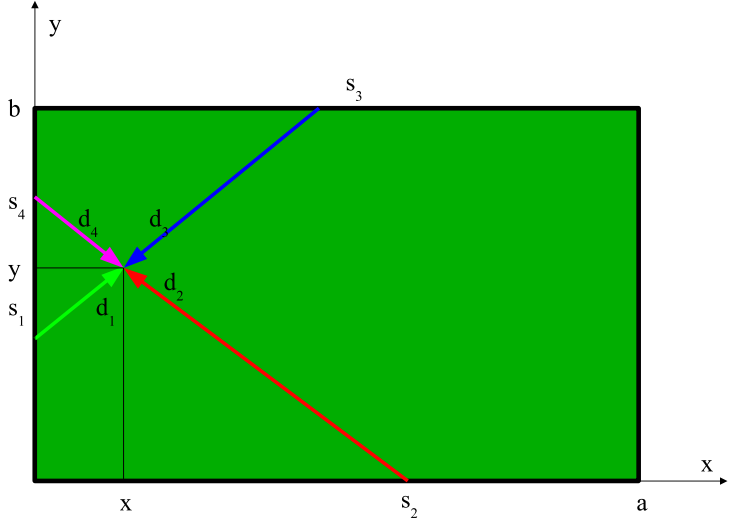,width=12cm}
\caption{The four rays of the four compound bundles and the corresponding four solutions $\Psi_i^+(d_i,s_i),\;i=1,...,4$, meeting at the point ($x,y$)}
\end{center}
\end{figure}

The above compound bundles are shown on Fig.3 where in each billiard point four their rays are met and the four
solutions \mref{A0b} are superposed to get GSWF, i.e.
\be
\Psi^{as}(x,y)=\sum_{k=1}^4{\tilde\Psi}_k^+(d_k,s_k,\lambda)
\label{A0d}
\ee

Since in our further considerations we will work exceptionally with the solutions \mref{A0b} we will drop the signature
of these solutions as well as the tilde mark for a convenience. Then assume the following standard forms for
$\Psi_k(d,s,\lambda),\;k=1,...,4,$ :
\be
\Psi_k(d,s,\lambda)=e^{i\lambda pd+i\lambda ps\cos\alpha_k}\chi_k(d,s,\lambda)
\label{A1b}
\ee
where $\alpha_k$ is the angle the momentum of the ray of the compound bundle $B_k$ makes with the corresponding side of the
rectangle measured anticlockwise.

The solution $\Psi_k(d,s,\lambda)$ is defined on the compound bundle $B_k$ which rays start from the sides $A_{k-1}$ and
$A_k$ so that the variable $s$ is measured from the left end of the corresponding side $A_k,\;k=1,...,4$.
For a given $\Psi_k(d,s,\lambda)$ $s$ is then positive on $A_k$ and negative on $A_{k-1}$ where $\Psi_k(d,s,\lambda)$
is also defined. Since the constant Jacobean factors have been included into $\chi$-coefficients the coefficient
$\chi_k(d,s,\lambda)$ is continuous on the sides $A_{k-1}\cup A_k,\;k=1,...,4$.

Let the colours of rays corresponding to the particular compound bundles  $B_k$ denote also colours of these bundles. Then
unfolding the skeleton of Fig.3 onto the plane a motion of the billiards ball which begins with the rays of the bundle
$B_1$ is limited by the stripe bounded by the two parallel (thick black) lines shown in Fig.4 and are performed along
the straight line. It is seen clearly on the figure that this motion is just the
scattering of the skeleton bundles on the (white) vertices of the rectangle so that each bundle is scattered into the two
neighbour ones with the exception of its single ray which crosses the vertex. The latter ray is scattered back into
the third remaining bundle. No one of the straight line rays crosses the rectangle boundary at the same point and
the crossing points of each ray are dense on each rectangle side. The ray of Fig.4 which starts at the point with
the coordinate $s$ on the figure is also shown in folded way on Fig.5.

The GSWF \mref{A0d} has to vanish on each side of the rectangle and this condition exhausts all the conditions it has to
satisfy. However, a particular form of the corresponding conditions depends on a choice of points on the rectangle
billiards boundary even for the same bundle. Therefore let us choose for writing these conditions the four first
points of the ray (including the starting point) shown in Fig.5. as convenient for our further considerations.  As it
follows from the form \mref{A0d} of the solution and from Fig.4 and Fig.5 the corresponding conditions are:
\be
e^{-i\lambda ps\sin\alpha}\chi_1(0,s,\lambda)+
e^{i\lambda p(a\cos\alpha-s\sin\alpha)}\chi_2\ll(\frac{-s}{\sin\alpha},-a-s\cot\alpha,\lambda\r)+\nn\\
e^{i\lambda p(a\cos\alpha+(b+s)\sin\alpha)}\chi_3\ll(\frac{b+s}{\sin\alpha},a-(b+s)\cot\alpha,\lambda\r)+\nn\\
e^{i\lambda p(b+s)\sin\alpha}\chi_4(0,b+s,\lambda)=0
\label{A1c}
\ee

\be
e^{i\lambda p(\frac{b}{\sin\alpha}+s\cot\alpha\cos\alpha)}\chi_1\ll(\frac{b+s}{\sin\alpha},s,\lambda\r)+\nn\\
e^{i\lambda p(b\sin\alpha+(a-(b+s)\cot\alpha)\cos\alpha)}\chi_2\ll(\frac{b}{\sin\alpha},(2b+s)\cot\alpha-a,\lambda\r)+\nn\\
e^{i\lambda p((a -(b+s)\cot\alpha)\cos\alpha)}\chi_3(0,a-(b+s)\cot\alpha,\lambda)+\nn\\
e^{i\lambda p(b+s)\cot\alpha\cos\alpha}\chi_4(0,-(b+s)\cot\alpha,\lambda)=0
\label{A1d}
\ee

\be
e^{i\lambda p(2b+s)\cot\alpha\cos\alpha}\chi_1(0,(2b+s)\cot\alpha,\lambda)+\nn\\
e^{i\lambda p(a\cos\alpha-(2b+s)\cot\alpha\cos\alpha)}\chi_2(0,(2b+s)\cot\alpha-a,\lambda)+\nn\\
e^{i\lambda p(a\cos\alpha+b\sin\alpha-(2b+s)\cot\alpha\cos\alpha)}\chi_3\ll(\frac{a-(2b+s)\cot\alpha}{\cos\alpha},3b+s-a\tan\alpha,\lambda\r)+\nn\\
e^{i\lambda p(\frac{b}{\sin\alpha}+(b+s)\cot\alpha\cos\alpha)}\chi_4\ll(\frac{b}{\sin\alpha},-(b+s)\cot\alpha,\lambda\r)=0
\label{A1e}
\ee

\be
e^{i\lambda p(\frac{a}{\cos\alpha}-(2b+s)\sin\alpha)}\chi_1\ll(\frac{a-(2b+s)\cot\alpha}{\cos\alpha},(2b+s)\cot\alpha,\lambda\r)+\nn\\
e^{i\lambda p(a\tan\alpha\sin\alpha-(2b+s)\sin\alpha)}\chi_2(0,a\tan\alpha-2b-s,\lambda)+\nn\\
e^{i\lambda p(-a\tan\alpha\sin\alpha +(3b+s)\sin\alpha)}\chi_3(0,3b+s-a\tan\alpha,\lambda)+\nn\\
e^{i\lambda p((3b+s)\sin\alpha-a\tan\alpha\sin\alpha+a\cos\alpha)}\chi_4\ll(\frac{3b+s-a\tan\alpha}{\sin\alpha},(3b+s)\cot\alpha-2a,\lambda\r)=0
\label{A1f}
\ee
where according to our convention $s$ is negative being measured from the left end of the side $A_1$ of the rectangle.

\begin{figure}
\begin{center}
\psfig{figure=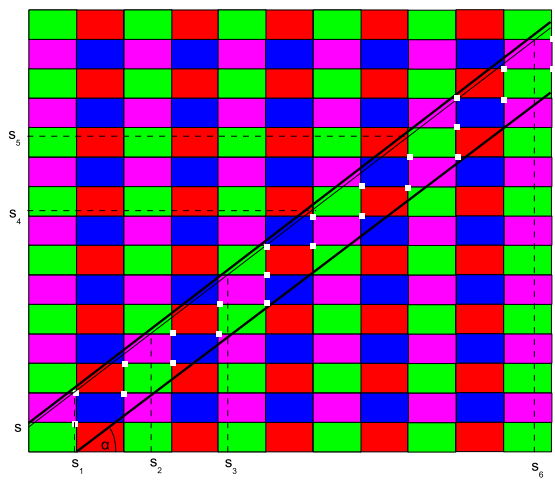,width=9cm}
\caption{The unfolded motion in the rectangular billiards}
\end{center}
\end{figure}

\begin{figure}
\begin{center}
\psfig{figure=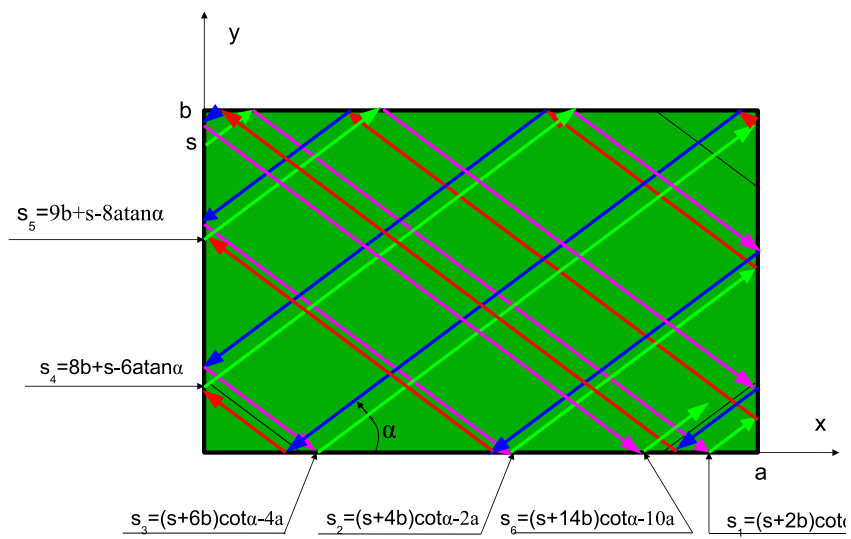,width=12cm}
\caption{The solution $\Psi_1(d,s)$ being carried by the successive bundles
$B_1\to B_4\to B_1\to B_2\to B_3\to B_4\to B_1\to ...$ . The corresponding ray is shown also unfolded in Fig.4}
\end{center}
\end{figure}

The last equations reduce to the following ones:
\be
\chi_1(0,s,\lambda)+e^{i\lambda pa\cos\alpha}\chi_2\ll(\frac{-s}{\sin\alpha},-a-s\cot\alpha,\lambda\r)=0\nn\\
e^{i\lambda pa\cos\alpha}\chi_3\ll(\frac{b+s}{\sin\alpha},a-(b+s)\cot\alpha,\lambda\r)+\chi_4(0,b+s,\lambda)=0
\label{A1}
\ee

\vskip 25pt

\be
e^{i\lambda pb\sin\alpha}\chi_1\ll(\frac{b+s}{\sin\alpha},s,\lambda\r)+\chi_4(0,-(b+s)\cot\alpha,\lambda)=0\nn\\
e^{i\lambda pb\sin\alpha}\chi_2\ll(\frac{b}{\sin\alpha},(2b+s)\cot\alpha-a,\lambda\r)+
\chi_3(0,a-(b+s)\cot\alpha,\lambda)=0
\label{A2}
\ee

\be
\chi_1(0,(2b+s)\cot\alpha,\lambda)+e^{i\lambda pb\sin\alpha}\chi_4\ll(\frac{b}{\sin\alpha},-(b+s)\cot\alpha,\lambda\r)=0\nn\\
\chi_2(0,(2b+s)\cot\alpha-a,\lambda)+
e^{i\lambda pb\sin\alpha}\chi_3\ll(\frac{a-(2b+s)\cot\alpha}{\cos\alpha},3b+s-a\tan\alpha,\lambda\r)=0
\label{A3}
\ee

\be
e^{i\lambda pa\cos\alpha}\chi_1\ll(\frac{a-(2b+s)\cot\alpha}{\cos\alpha},(2b+s)\cot\alpha,\lambda\r)+
\chi_2(0,a\tan\alpha-2b-s,\lambda)=0\nn\\
\chi_3(0,3b+s-a\tan\alpha,\lambda)+
e^{i\lambda pa\cos\alpha}\chi_4\ll(\frac{3b+s-a\tan\alpha}{\sin\alpha},(3b+s)\cot\alpha-2a,\lambda\r)=0
\label{A4}
\ee

The first two equations of \mref{A2} and \mref{A3} can be "solved" in the first order in $\lambda$ with the help of
\mref{12z} to get:
\be
\chi_{1,0}((2b+s)\cot\alpha)=e^{2i\lambda pb\sin\alpha}\chi_{1,0}(s)
\label{A5}
\ee

The point $s=(2b+s)\cot\alpha$ lies now on the side $A_1$ of the rectangle. It is achieved by the ray leaving the
starting point $s$ of the side $A_4$ of the rectangle and passing the distance $\frac{2b+s}{\sin\alpha}$. Since the
coefficient $\chi_{1,0}(s)$ does not change propagating along the ray we have to have:
\be
\chi_{1,0}((2b+s)\cot\alpha)=\chi_{1,0}(s)
\label{A6}
\ee
so that
\be
e^{2i\lambda pb\sin\alpha}=1
\label{A7}
\ee
and
\be
\lambda pb\sin\alpha=n\pi,\;\;\;\;\;\;n=1,2,....
\label{A8}
\ee

Similarly, choosing another propagation path we get:
\be
e^{2i\lambda pa\cos\alpha}=1
\label{A9}
\ee
and
\be
\lambda pa\cos\alpha=m\pi,\;\;\;\;\;\;m=1,2,....
\label{A10}
\ee

On the other hand the equation \mref{A6} is of great importance because it shows that $\chi_{1,0}(s)$ is defined
in different points of the sides $A_1,\;A_4$ by its value established in some definite point of these sides and by
the propagation procedure defined by  \mref{12} and \mref{12z}. Since however a propagated ray reflects consecutively
on the boundary in points densely distributed on it this initial value is also propagated densely on the boundary.
Therefore to get the coefficient as a continuous function of $s$ we have to put it a constant. Let it be equal to one.

The next order term propagates according to the formula:
\be
\chi_{1,1}(d,s)=\chi_{1,1}(s)+\frac{iE_1}{p}d
\label{A11}
\ee
where $\chi_{1,1}(s)$ is the initial value of the term on the boundary. Therefore, since distances $d_i$ measured
along the propagating ray of the consecutive boundary points $s_i$ by which the ray is reflected are distributed on the
boundary irregularly but densely values of $\chi_{1,1}(0,s_i)\equiv\chi_{1,1}(s_i)=\chi_{1,1}(d_i,s)$ in these points
have to change discontinuously. Therefore to maintain the continuity property of $\chi_{1,1}(s)$ on the boundary we have
to put $E_1=0$ in \mref{A11} so that $\chi_{1,1}(d,s)$ is again constant on the boundary and in consequence
independent also of $d$.

Quite similarly we can
argue that also the remaining terms have to be constant as well as all the other terms of the energy semiclassical
series have to vanish.

In this way we get finally from \mref{A2}-\mref{A5}:
\be
\chi_{1,mn}(d,s,\lambda)\equiv 1\nn\\
\chi_{2,mn}(d,s,\lambda)\equiv (-1)^{m+1}\nn\\
\chi_{3,mn}(d,s,\lambda)\equiv (-1)^{m+n}\nn\\
\chi_4(d,s,\lambda)\equiv (-1)^{n+1}
\label{A12}
\ee
and
\be
E_{mn}=E_{0,mn}=\fr p_{mn}^2=\frac{\pi^2}{2\lambda^2}\ll(\frac{m^2}{a^2}+\frac{n^2}{b^2}\r),\;\;\;\;\;\;m,n=1,2,3,...
\label{A13}
\ee

Choosing therefore the point $(x,y)$ of Fig.3 for the SWF \mref{A0d} we get:
\be
\Psi_{mn}^{as}(x,y)=\sum_{l=1}^4\Psi_l(d_l,s_l)=
\sum_{l=1}^4e^{i\lambda p_{mn}d_l+i\lambda p_{mn}f_l(s_l)+\phi_l}
\label{A17}
\ee
where $d_l$ and $f_l(s_l),\;l=1,...,4$ should be calculated from the relations (see Fig.5):
\be
x=x_0(s_l)+d_l\cos\alpha(s_l)\nn\\
y=y_0(s_l)+d_l\sin\alpha(s_l)\nn\\
\alpha(s_l)=\alpha,\;\fr\pi-\alpha
\label{A18}
\ee
and the phases $\phi_l$ are defined by \mref{A12}.

However making use of the independence of the phase integral $\int_{(0,0)}^{(x,y)}p_xdx+p_ydy$ of the integration
paths we get for the particular terms in the sum in \mref{A17}:
\be
e^{i\lambda pd_1+i\lambda pf_1(s_1)}=e^{i\lambda(p_xx+p_yy)}\nn\\
e^{i\lambda pd_2+i\lambda pf_2(s_2)+i(n+1)\pi}=e^{i\lambda(p_x(a-x)+p_yy+(n+1)\pi)}=-e^{i\lambda(-p_xx+p_yy)}\nn\\
e^{i\lambda pd_3+i\lambda pf_3(s_3)+i(m+n)\pi}=e^{i\lambda(p_x(a-x)+p_y(b-y)+(m+n)\pi)}=e^{i\lambda(-p_xx-p_yy)}\nn\\
e^{i\lambda pd_4+i\lambda pf_4(s_4)+i(m+1)}=e^{i\lambda(p_xx+p_y(b-y)+(m+1)\pi)}=-e^{i\lambda(p_xx-p_yy)}
\label{A19}
\ee
where $p_x=p_{mn}\cos\alpha_{mn},\;p_=p_{mn}\sin\alpha_{mn}$.

Finally:
\be
\Psi_{mn}^{as}(x,y)=e^{i\lambda(p_xx+p_yy)}-e^{i\lambda(-p_xx+p_yy)}+e^{i\lambda(-p_xx-p_yy)}-e^{i\lambda(p_xx-p_yy)}=\nn\\
-4\sin(p_xx)\sin(p_yy),\;\;\;\;\;\;m,n=1,2,3,...
\label{A20}
\ee
reproducing in this way the exact result \mref{A0a}.

Let us note that the set of all SWF's \mref{A20} is again complete.

Of course it is not surprising that the semiclassical calculations performed above reproduce the exact result \mref{A0a}
since it follows from its form that it represents simultaneously its semiclassical expansion. However to get the
result \mref{A20} we have had to assume that skeletons considered had to be "generic", i.e. they were constructed of
non-periodic trajectories. Just this assumption allows us to use the arguments of dens distributions of values of the SWF to
establish its value on the boundary. In fact no other argument exists to get such a conclusion. On the other hand
such an argument cannot be invoked in the cases of skeletons which are built of periodic trajectories so that such
cases of skeletons must be considered separately.

Finally let us conclude that:
\begin{enumerate}
\item each non-periodic skeleton in the rectangular billiards is regular and equivalent in the phase space to a two
dimensional torus;
\item the GSWF \mref{A20} is obviously regular and provides us with the exact solution to the SE;
\item the GSWF \mref{A20} as well as the corresponding energy levels coincide with their JWKB approximations since
the corresponding semiclassical series for the SWF's and energy levels abbreviate on the zeroth term; and
\item it is the dense distribution of the skeleton rays in the configuration and the phase spaces which causes the
semiclassical series abbreviation mentioned.
\end{enumerate}

\subsection{Skeletons built by periodic trajectories}

\hskip+2em Periodic skeletons in the rectangular billiards can be easily realized since bundles of such
trajectories with a given period are defined by one of its members which starting from a vertex of the rectangle has to be
reflected in other vertices to "finish" its motion in the initial vertex. Such a leading particular
periodic trajectory will be called a singular diagonal (SD) after Bogomolny and Schmit \cite{1}. All other trajectories which make
the same angles with the corresponding sides of the rectangle as this particular SD does are then also periodic.

\begin{figure}
\begin{center}
\psfig{figure=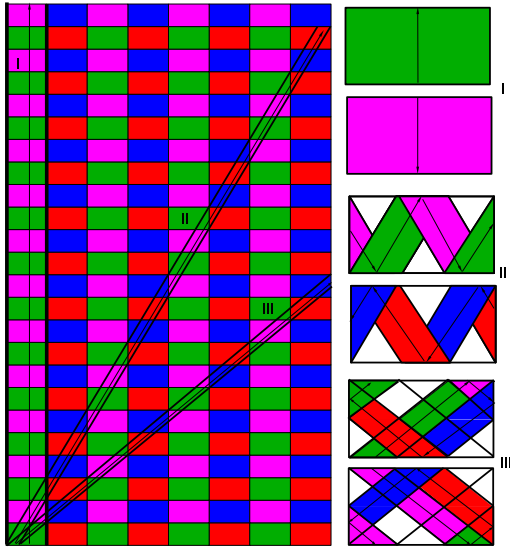,width=7cm}
\caption{Three unfolded skeletons defined by the corresponding pairs of SD's. The skeletons $II$ and $III$ are
singular. Every skeleton is the cylinder-like Lagrange surface in the phase space.}
\end{center}
\end{figure}

In the rational billiards periodic skeletons are defined always by two such SD's. In the rectangular billiard these
two SD's
are symmetric with respect to each other in a sense that if one of them crosses some two different vertices of the
rectangle the second one has to cross the remaining two.

On Fig.6 and Fig.7 there are shown five cases of skeletons defined by pairs of such SD's represented in their unfolded form
(the left picture) and in their real form in the
billiards (the right pictures). Each pair of SD's defining each skeleton are visible in the unfolded form of the skeletons
as two parallel straight lines. Each skeleton is a stripe bounded by such two SD's. A general property of each such a
stripe is that all the rectangle vertices related with the stripe lie on its boundary, i.e. on its two SD's.

Single periodic trajectories
are shown also in each skeleton case in the figure being parallel to the SD's defining skeletons. In the billiards
(the right pictures) these periodic trajectories are of course closed. The skeletons on the figure have forms which are
typical, i.e. infinitely many others differ from these on the figure by a number of reflections of SD's on the
rectangle sizes.

\subsubsection{Bouncing ball skeleton and the corresponding regular GSWF's}

\hskip+2em We will construct GSWF's on these skeletons with the same rules as formulated earlier. Let us begin with the case
of the skeleton numbered by $I$ in Fig.6 and shown in Fig.8. This bouncing ball skeleton contains only two bundles
$B_1$ and $B_3$ -
the first one with its rays directed up and starting from the side $A_1$ and the second $B_3$ with rays directed down
starting from the side $A_3$. The skeletons ${\bf B}^A$ and ${\bf B}^R$ are identical with ${\bf B}$ which is of course
regular.

\begin{figure}
\begin{center}
\psfig{figure=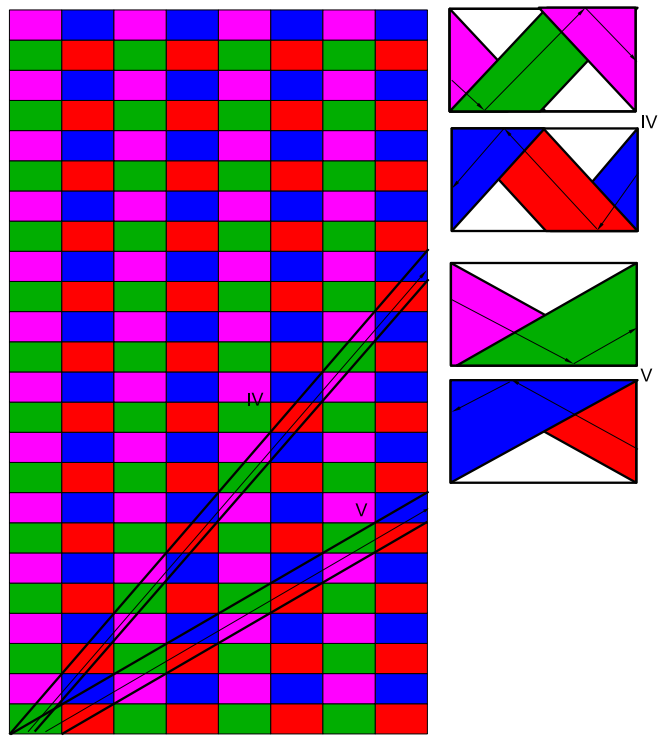,width=9cm}
\caption{Another two unfolded singular skeletons defined by the corresponding pairs of SD's with the same cylinder-like
Lagrange surfaces in the phase space.}
\end{center}
\end{figure}

For these particular cases of bundles rays for both the bundles will be positioned by the same parameter $s$ measuring
a distance of a ray from the $y$-axis along the corresponding sides $A_1$ and $A_3$. Therefore for the corresponding
BSWF's:
\be
\Psi_1(d,s,\lambda)=e^{\lambda pd}\chi_1(d,s,\lambda)\nn\\
\Psi_3(b-d,s,\lambda)=e^{i\lambda p(b-d)}\chi_3(b-d,s,\lambda)\nn\\
0\leq d\leq b,\;\;\;\;0<s<a
\label{A21}
\ee

For the coefficients $\chi_k(d,s,\lambda),\;k=1,3$ it is assumed as usually that they propagate along the rays of
the bundles continuously and this their property is not influenced by reflections of the rays on the boundaries.
Therefore we have to accept also that they are periodic with respect to the $d$-variable with the period equal to $2b$.

For the GSWF $\Psi^{as}(x,y,\lambda)$ we have therefore:
\be
\Psi^{as}(x,y,\lambda)=\Psi_1(y,x,\lambda)+\Psi_3(b-y,x,\lambda)
\label{A22}
\ee
together with the following boundary conditions on the sides $A_1$ and $A_3$ respectively:
\be
\chi_1(0,x,\lambda)+e^{i\lambda pb}\chi_3(b,x,\lambda)=0\nn\\
e^{i\lambda pb}\chi_1(b,x,\lambda)+\chi_3(0,x,\lambda)=0
\label{A23}
\ee
so that:
\be
\chi_{1,0}(x)=e^{2i\lambda pb}\chi_{1,0}(x)
\label{A24}
\ee

As previously we conclude that:
\be
\lambda pb=n\pi,\;\;\;\;\;\;n=1,2,3,...
\label{A25}
\ee

\begin{figure}
\begin{center}
\psfig{figure=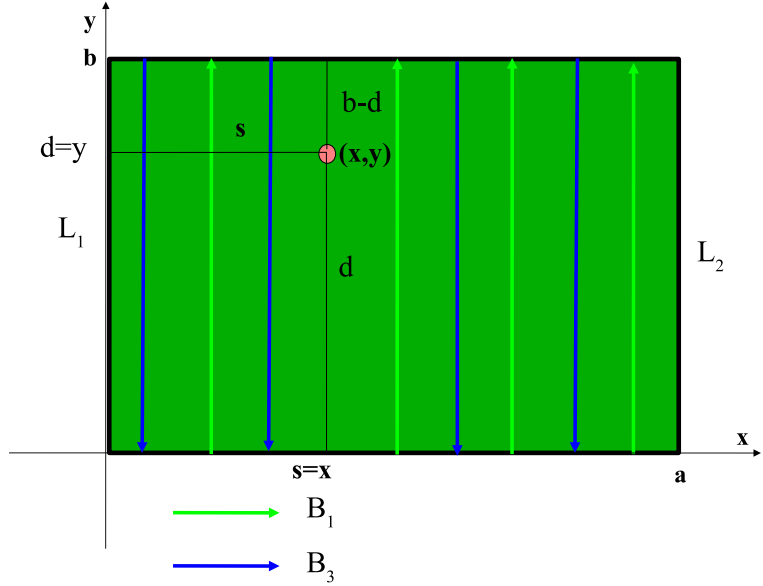,width=12cm}
\caption{The two bouncing mode bundles $B_1$ and $B_3$ of the regular vertical skeleton in the rectangular billiards}
\end{center}
\end{figure}

Now the corresponding boundary conditions for $\Psi_k(y,x,\lambda),\;k=1,3$ on the sides $A_2$ and $A_4$ give:
\be
\chi_k(y,0,\lambda)=\chi_k(y,a,\lambda)\equiv 0\nn\\
k=1,3
\label{A29}
\ee

Therefore in the zeroth order we have:
\be
\chi_{1,0}(0)=\chi_{1,0}(a)=0
\label{A30}
\ee

Next let us invoke the second of the equations \mref{12z} and the periodicity of $\chi_1(y,x,\lambda)$ to get in the
considered case for the second order term:
\be
\chi_{1,1}(2b,x)=\chi_{1,1}(0,x)=\chi_{1,1}(0,x)+\frac{ib}{2p}\ll(\frac{d^2\chi_{1,0}(x)}{dx^2}+2E_1\chi_{1,0}(x)\r)
\label{A31}
\ee
so that
\be
\frac{d^2\chi_{1,0}(x)}{dx^2}+2E_1\chi_{1,0}(x)=0
\label{A32}
\ee

The obvious solution of the last equation satisfying the boundary conditions \mref{A29} is:
\be
\chi_{1,0}(x)=A_0\sin(\sqrt{2E_1}x)\nn\\
\sqrt{2E_1}a=m\pi,\;\;\;\;\;m=1,2,...
\label{A33}
\ee

Coming back to the second of the equations \mref{12z} we can conclude that $\chi_{1,1}(y,x)$ again is independent of
$y$.

Passing next to the third of the equations \mref{12z} and repeating arguments similar to those which led us to
\mref{A31} we get the following equation for $\chi_{1,1}(x)$:
\be
\frac{d^2\chi_{1,1}(x)}{dx^2}+2E_1\chi_{1,1}(x)+2E_2\chi_{1,0}(x)=0
\label{A32a}
\ee
with the solution:
\be
\chi_{1,1}(x)=A_1\sin(\sqrt{2E_1}x)+B_1\cos(\sqrt{2E_1}x)+\frac{E_2A_0x}{\sqrt{2E_1}}\cos(\sqrt{2E_1}x)
\label{A33}
\ee

The boundary conditions $\chi_{1,1}(0)=\chi_{1,1}(a)=0$ enforce however $B_1=E_2=0$.

Using again \mref{12z} and the inductive arguments we come to the conclusion that $\chi_1(y,x,\lambda)$ is
$y$-independent and the coefficients of its semiclassical series have the form:
\be
\chi_{1,k}(x)=A_k\sin(\sqrt{2E_1}x),\;\;\;\;\;\;k=0,1,...
\label{A34}
\ee
so is the form of $\chi_1(x,\lambda)$ itself, i.e.
\be
\chi_1(x,\lambda)=A(\lambda)\sin(\sqrt{2E_1}x)=A(\lambda)\sin\ll(m\pi\frac{x}{a}\r)\nn\\
A(\lambda)=\sum_{k\geq 0}A_k\lambda^{-k}
\label{A35}
\ee

Clearly, similar conclusion can be obtained for $\chi_3(y,x,\lambda)$ which by \mref{A23} and for the $mn$-th energy
level is equal to:
\be
\chi_{3,mn}(y,x,\lambda)=-(-1)^n\chi_{1,m}(x,\lambda)=-(-1)^nA(\lambda)\sin\ll(m\pi\frac{x}{a}\r)
\label{A26}
\ee

Therefore coming back to \mref{A22} we get:
\be
\Psi_{mn}^{as}(x,y,\lambda)=2iA(\lambda)\sin\ll(n\pi\frac{y}{b}\r)\sin\ll(m\pi\frac{x}{a}\r)
\label{A36}
\ee
which again is the result got in the previous section.

The energy $E$ is given however by the {\bf finite} semiclassical series:
\be
E=\fr p^2+\frac{E_1}{\lambda^2}=\fr\ll(\ll(\frac{n\pi}{\lambda b}\r)^2+\ll(\frac{m\pi}{\lambda a}\r)^2\r),\;\;\;\;\;m,n=1,2,...
\label{A37}
\ee

Let us stress the following main differences between the previous non-periodic case and the bouncing mode one despite the fact that
in both the cases the results obtained are the same.
\begin{enumerate}
\item The bouncing ball skeleton is represented in the phase space by a cylinder rather than by a closed
torus;
\item contrary to the non-periodic cases only one skeleton is sufficient in the bouncing mode case to get the whole
spectrum of the energy;
\item all terms of the semiclassical series expansion of GSWF exist (do not vanish) in the bouncing mode case, while only the zeroth one in
the non-periodic one;
\item the semiclassical series for the energy contains two first terms in the bouncing mode case and only zeroth
non-vanishing term in the non-periodic one; and
\item the JWKB approximation of the energy does {\bf not} coincide with its global value.
\end{enumerate}

However similarly to the non-periodic case the bouncing ball solution is also regular and exact.

\subsubsection{Periodic skeletons different than the bouncing ball ones - singular SWF's}

\hskip+2em Consider now skeletons which periodic rays do not bounce between the sides of the rectangle. The
simplest such a case the fifth one in Fig.7 is shown in Fig.9. As in the non-periodic case there are again four bundles
in the corresponding skeleton but contrary to the case mentioned only two rays (of four of them) belonging to two
different bundles can meet at each point of the rectangle if this point does not lie on SD's.

We have to note also that the skeleton associated with this on Fig.9 differs from it by the opposite directions of rays,
i.e. possible energy levels we get for these skeletons must be degenerated.

The GSWF corresponding to the case looks as follows in different domains of the rectangles:
\be
\Psi^{as}(x,y)=\ll\{\ba{lr}
		e^{i\lambda pd_1+i\lambda ps_1\cos\alpha}\chi_1(d_1,s_1,\lambda)+
                e^{i\lambda pd_4'+i\lambda ps_4'\sin\alpha}\chi_4(d_4',s_4',\lambda)&(x,y)\in D_1\\
                e^{i\lambda pd_1'+i\lambda ps_1'\cos\alpha}\chi_1(d_1',s_1',\lambda)+
                e^{i\lambda pd_2+i\lambda ps_2\sin\alpha}\chi_2(d_2,s_2,\lambda)&(x,y)\in D_2\\
                e^{i\lambda pd_3+i\lambda ps_3\cos\alpha}\chi_3(d_3,s_3,\lambda)+
                e^{i\lambda pd_2'+i\lambda ps_2'\sin\alpha}\chi_2(d_2',s_2',\lambda)&(x,y)\in D_3\\
                e^{i\lambda pd_3'+i\lambda ps_3'\cos\alpha}\chi_3(d_3',s_3',\lambda)+
                e^{i\lambda pd_4+i\lambda ps_4\sin\alpha}\chi_4(d_4,s_4,\lambda)&(x,y)\in D_4\\
                \tan\alpha=\frac{b}{a}&
                                \ea\r.
                \label{A38}
\ee
where the variables $s_k,s_k'$ are measured from the left ends of the corresponding sides $A_k,\;k=1,...,4$.

The Dirichlet boundary conditions on the respective sides of the rectangle are therefore:
\be
\chi_1(0,s,\lambda)+e^{i\lambda pb\sin\alpha}\chi_4\ll(\frac{s}{\cos\alpha},b-s\tan\alpha,\lambda\r)=0\nn\\
e^{i\lambda pa\cos\alpha}\chi_1\ll(\frac{a-s}{\cos\alpha},s,\lambda\r)+\chi_2(0,(a-s)\tan\alpha,\lambda)=0\nn\\
\chi_3(0,s,\lambda)+e^{i\lambda pb\sin\alpha}\chi_2\ll(\frac{s}{\cos\alpha},(a-s)\tan\alpha,\lambda\r)=0\nn\\
e^{i\lambda pa\cos\alpha}\chi_3\ll(\frac{a-s}{\cos\alpha},s,\lambda\r)+\chi_4(0,b-s\tan\alpha,\lambda)=0\nn\\
0<s<a
\label{A39}
\ee

\begin{figure}
\begin{center}
\psfig{figure=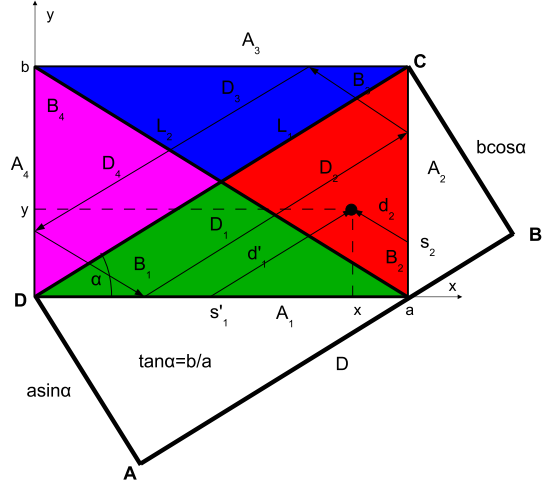,width=9cm}
\caption{The singular skeleton defined by the corresponding pair $L_1$ and $L_2$ of SD's}
\end{center}
\end{figure}

One can easily find from \mref{A39} that:
\be
\chi_1(0,s,\lambda)=e^{2i\lambda p(b\sin\alpha+a\cos\alpha)}\chi_1\ll(\frac{2a}{\cos\alpha},s,\lambda\r)
\label{A40}
\ee
or
\be
e^{2i\lambda p(b\sin\alpha+a\cos\alpha)}=1
\label{A41}
\ee
because $\chi_1(d,s,\lambda)$ is periodic with the period $\frac{2a}{\cos\alpha}=2(b\sin\alpha+a\cos\alpha)=2D$
where $D$ is the length of the rectangle diagonal.

Therefore we get the following quantization condition for the zeroth energy term $E_0=\fr p^2$
\be
\lambda pD=n\pi,\;\;\;\;\;\;\;\;n=1,2,...
\label{A42}
\ee

Now we have to note that none of the bundle considered has a piece of its boundary common with any other one inside the
rectangle. Seemingly as
such could be considered the rectangle diagonals if the rays in the respective bundles were not run in the opposite
directions. Therefore GSWF's defined in these bundles have to vanish on the respective diagonals of the rectangle so
that we have to have:
\be
\chi_k(d,0,\lambda)=0,\;\;\;\;\;\;k=1,...,4\nn\\
0\leq d\leq D
\label{A44}
\ee

But then from \mref{A39} we get also:
\be
\chi_1(0,a,\lambda)=\chi_2(0,b,\lambda)=\chi_3(0,a,\lambda)=\chi_4(0,b,\lambda)=0
\label{A45}
\ee

Further using the propagation formula \mref{12z} for $\chi_{1,1}(2D,s)=\chi_{1,1}(0,s)$ we get:
\be
\chi_{1,0}''(s)+2E_1\sin^2\alpha\chi_{1,0}(s)=0
\label{A45a}
\ee
which with the conditions \mref{A44}-\mref{A45} for $\chi_1(d,s,\lambda)$ gives:
\be
\chi_{1,0}(s)=A_0\sin(\sqrt{2E_1}\sin\alpha s)
\label{A46}
\ee
where $E_1$ defines the second term of the semiclassical energy expansion with the condition:
\be
E_1=\fr\frac{m^2\pi^2}{a^2\sin^2\alpha},\;\;\;\;\;\;\;m=1,2,3,...
\label{A47}
\ee

Next repeating the procedure for the bouncing mode skeleton to the remaining terms
$\chi_{1,k}(2D,s)=\chi_{1,k}(0,s),\;k=1,2,3,...,$
we get for them:
\be
\chi_{1,k}(s)=A_k\sin(\sqrt{2E_1}\sin\alpha s)
\label{A46a}
\ee
so that
\be
\chi_1(d,s,\lambda)=A(\lambda)\sin(\sqrt{2E_1}\sin\alpha s)\nn\\
A(\lambda)=\sum_{k\geq 0}A_k\lambda^{-k}
\label{A47a}
\ee

Therefore using \mref{A39} the final form of the SWF $\Psi^{as}(x,y)$ can be written as follows:
\be
\Psi_{mn}^{as}(x,y)=\ll\{\ba{lr}
               e^{i\lambda p_n(x\cos\alpha+y\sin\alpha)}\sin\ll(m\pi\frac{1}{a}(x-y\cot\alpha)\r)-\\
               e^{i\lambda p_n(x\cos\alpha-y\sin\alpha)}\sin\ll(m\pi\frac{1}{a}(x+y\cot\alpha)\r)&(x,y)\in D_1\\
               e^{i\lambda p_n(x\cos\alpha+y\sin\alpha)}\sin\ll(m\pi\frac{1}{a}(x-y\cot\alpha)\r)+\\
               e^{i\lambda p_n(-x\cos\alpha+y\sin\alpha+2a\cos\alpha)}\sin\ll(m\pi\frac{1}{a}(x+y\cot\alpha)\r)
               &(x,y)\in D_2\\
               -e^{i\lambda p_n(-x\cos\alpha-y\sin\alpha)}\sin\ll(m\pi\frac{1}{a}(x-y\cot\alpha)\r)+\\
               e^{i\lambda p_n(-x\cos\alpha+y\sin\alpha+2a\cos\alpha)}\sin\ll(m\pi\frac{1}{a}(x+y\cot\alpha)\r)
                &(x,y)\in D_3\\
                -e^{i\lambda p_n(-x\cos\alpha-y\sin\alpha)}\sin\ll(m\pi\frac{1}{a}(x-y\cot\alpha)\r)-\\
                e^{i\lambda p_n(x\cos\alpha-y\sin\alpha)}\sin\ll(m\pi\frac{1}{a}(x+y\cot\alpha)\r)&(x,y)\in D_4
                \ea\r.
\label{A48}
\ee
while the energy spectrum is:
\be
E=\fr p^2+\frac{E_1}{\lambda^2}=\frac{\pi^2}{2\lambda^2}\ll(\frac{n^2}{D^2}+\frac{m^2D^2}{a^2b^2}\r),
\;\;\;\;\;\;\;m,n=1,2,3,...
\label{A49}
\ee

By their construction the solutions \mref{A48} are all singular - their derivatives are discontinuous on the rectangle diagonals.
The corresponding energy spectrum also differs from the exact one as well as from the one of the regular SWF's obtained
in the last two sections.

It should be stressed however that the solutions \mref{A48} - \mref{A49} are {\bf allowed} semiclassical solutions to the rectangle
billiards eigenvalue problem, which corresponds physically to effects of the short wave limits. It is therefore of
great importance whether one can detect in these limits resonant modes in the respective rectangular cavity corresponding
to the SWF's \mref{A48} and the energy spectrum \mref{A49}. Indeed
such modes have been detected experimentally by Bogomolny {\it et al} \cite{2} for the rectangle cavity with a barrier
inside. This case of the billiards will be discussed in the next sections.

If however such modes can be detected in the rectangle cavity then one can expect the corresponding GSWF's to have forms
of standing waves rather than of the running ones as in \mref{A48}. We can get such forms of GSWF's noticing that
the spectrum \mref{A49} is obviously degenerate since in the case considered the associated skeleton ${\bf B}^T$ is
different from {\bf B}. We can use therefore the corresponding running solutions for the skeleton ${\bf B}^T$ to
construct by superpositions the standing SWF's corresponding to the energy spectrum \mref{A49}. The simplest two
superpositions are:
\be
\Psi_{1,2;mn}^{as}(x,y)=\nn\\\ll\{\ba{lr}
	        \sin\ll(m\pi\frac{1}{a}(x-y\cot\alpha)\r)\times
	        \ll\{\ba{l}\sin(\lambda p_n(x\cos\alpha+y\sin\alpha))\\
	        \pm\cos(\lambda p_n(x\cos\alpha+y\sin\alpha))
	        \ea\r.-\\
	         \sin\ll(m\pi\frac{1}{a}(x+y\cot\alpha)\r)\times
	          \ll\{\ba{l}\sin(\lambda p_n(x\cos\alpha-y\sin\alpha))\\
	        \cos(\lambda p_n(x\cos\alpha-y\sin\alpha))
	        \ea\r.&(x,y)\in D_1,\;D_4\\&\\
	        \sin\ll(m\pi\frac{1}{a}(x-y\cot\alpha)\r)\times
	        \ll\{\ba{l}\sin(\lambda p_n(x\cos\alpha+y\sin\alpha))\\
	        \pm\cos(\lambda p_n(x\cos\alpha+y\sin\alpha))
	        \ea\r.-\\
	         \sin\ll(m\pi\frac{1}{a}(x+y\cot\alpha)\r)\times
	          \ll\{\ba{l}\sin(\lambda p_n(x\cos\alpha-y\sin\alpha-2a\cos\alpha))\\
	        -\cos(\lambda p_n(x\cos\alpha-y\sin\alpha-2a\cos\alpha))
	        \ea\r.&(x,y)\in D_2,\;D_3
	        \ea\r.
\label{A50}
\ee
where the plus corresponds to the domains $D_1,D_2$ and the minus - to $D_3,D_4$.

The above results on the singular SWF's can be generalized to arbitrary periodic skeletons in the rectangular billiards
some of which are shown on Fig.6 and Fig.7.
This can be done by noticing that an arbitrary periodic SD is defined by arbitrary two relatively prime numbers
$\{p,q\}$ so that a SD in a rectangle with the sides
$a$ and $b$ shown in Fig.9 starting from the vertex $(0,0)$ and being inclined by an angle $\alpha$ to the $x$-axis is
defined by such two numbers as follows:
\be
\tan\alpha=\frac{pb}{qa}
\label{A58}
\ee

The above fact follows directly from the unfolded forms of periodic trajectories shown in Fig.6 and Fig.7 if one
realizes that each of them has to finish on another vertex of the rectangle. It follows also that the set of all SD is
countable but dense among all trajectories in the rectangle.

Pairs $\{p,q\}$ can appear in the following combinations $\{e,o\}$, $\{o,o\}$ and $\{o,e\}$ where $e$ stands for "even"
and $o$ - for "odd".
The respective SD's defined by these combinations finish their runs in the vertices $(a,0)$, $(a,b)$ and $(0,b)$
correspondingly.

A SD defined by a pair $\{p,q\}$ bounces $p-1$-times from each horizontal side of the rectangle and $q-1$-times - from
each of the vertical ones. If $D$ denotes its global length measured from its starting vertex $(0,0)$ to one of its final
ones just enumerated then $D=pb\sin\alpha+qa\cos\alpha=\sqrt{(qa)^2+(pb)^2}$.

If a SD is chosen, i.e. $\{p,q\}$ are fixed, and it ends at one of the vertices just enumerated then the second SD which
has to accompany the chosen one
to built the skeleton starts and ends at the remaining two of these vertices. Note that a number of bundles in such a
skeleton is then equal to $2p+2q$ while their widths are equal to $\frac{a}{p}\sin\alpha=\frac{b}{q}\cos\alpha$.

The quantization formula \mref{A42} remains then valid for the case considered while \mref{A47} takes the form
$E_1=\fr\frac{m^2p^2\pi^2}{a^2\sin^2\alpha},\;m=1,2,3,...$, so that
the formula \mref{A49} for the energy spectrum remains also unchanged. This formula corresponds to the spectrum of
the exact standing wave functions in a rectangle with the sides $D\times\frac{a}{p}\sin\alpha$. According to Bogomolny
and Schmit \cite{1}
this rectangle can be considered as an unfolded skeleton so that each such a standing wave function defined on it
should generate a corresponding semiclassical one defined on the skeleton by folding appropriately the
rectangle mentioned to the real skeleton and interfering pieces of the standing wave function in crossed points of such
a folding. However such a procedure to be correct still needs for the resulting SWF's to vanish on the
rectangular billiards sides. It is seen that such a procedure though theoretically possible and correct is complicated
enough to be replaced by the
corresponding constructions of the SWF's in the real folded skeleton according to the rules formulated in sec.3.

Nevertheless the SWF's corresponding to the energy spectrum \mref{A49} are all singular having discontinuous derivatives
on the lines separating two neighboring bundles. The lines are just the bundles boundaries on which the running SWF's
defined in these bundles have to vanish.

\subsubsection{Regular SWF's in the periodic skeletons in rectangular billiards}

\hskip+2em The singular SWF's found in the previous section provides us with the energy spectrum which is different from
the regular one. However for a particular relations between the rectangle sides these solutions can become regular. This
can happen if one assumes the following two additional conditions:
\be
\lambda pb\sin\alpha=k\pi\nn\\
\lambda pa\cos\alpha=l\pi\nn\\
k,l=1,2,....
\label{A51}
\ee
where $p$ is the global momentum of the billiards ball.

While it is tedious to be checked the singular SWF's satisfying the conditions \mref{A51} become then regular obtaining
the following form:

\be
{\tilde\Psi}_{1,2;mn}^{as}(x,y)=
\sin\ll(\frac{m-l}{a}p\pi x\r)\sin\ll(\frac{m+k}{b}q\pi y\r)\pm
\sin\ll(\frac{m+l}{a}p\pi x\r)\sin\ll(\frac{m-k}{b}q\pi y\r)
\label{A52}
\ee
valid in the whole rectangle area.

An immediate consequence of the conditions \mref{A51} is that they limit the form of the rectangles for which the
solution \mref{A52} can exist. Namely we have to have:
\be
\tan^2\alpha=\ll(\frac{pb}{qa}\r)^2=\frac{k}{l}\;\;\;\;\;\;k,l=1,2,....
\label{A53}
\ee

Let us make further an important note that despite its semiclassical origin the solution \mref{A52} is still exact being
a linear combination of two solutions given by the formula \mref{A0a}. But since it is
an eigenfunction of a given energy this combination means that both its terms have to be eigenfunctions of the same
energy, i.e. we should have:
\be
\frac{(m-l)^2p^2}{a^2}+\frac{(m+k)^2q^2}{b^2}=\frac{(m+l)^2p^2}{a^2}+\frac{(m-k)^2q^2}{b^2}
\label{A54}
\ee

It is easy to check that this is the case if one takes into account the condition $(pb)^2l=(qa)^2k$ which follows from
\mref{A53}.

Therefore if $k_0,\;l_0$ for which $\alpha$ satisfies \mref{A53} are relatively prime then the remaining allowed
pairs of $k,\;l$ satisfying the condition \mref{A53} are of course of the form $k=nk_0,\;l=nl_0,\;n=1,2,...,$ and the energy
spectrum formula \mref{A49} for such a rectangle takes the following final form:
\be
E=\frac{\pi^2}{2\lambda^2}\frac{(m^2+n^2k_0l_0)(k_0+l_0)p^2}{a^2k_0},
\;\;\;\;\;\;\;m,n=1,2,3,...
\label{A56}
\ee

Each energy level \mref{A56} is degenerate with the following base in their two dimensional degeneracy space:
\be
\Phi_{1,mn}(x,y)=\sin\ll(\frac{m-nl_0}{a}p\pi x\r)\sin\ll(\frac{m+nk_0}{b}q\pi y\r)\nn\\
\Phi_{2,mn}(x,y)=\sin\ll(\frac{m+nl_0}{a}p\pi x\r)\sin\ll(\frac{m-nk_0}{b}q\pi y\r)\nn\\
m,n=1,2,3,...
\label{A57}
\ee

To conclude for the "periodic" skeleton considered in this section to get regular SWF's it is necessary for a rectangle
to satisfy first
the constrain \mref{A51}. But if it happens then the corresponding
SWF's coincide with the ones built on "generic" skeletons considered
in sec. 4.1. An additional result of the considered periodic skeleton configurations is that the corresponding energy
levels are degenerate.

Let us stress however that if there are no such four natural numbers $k,l,p,q$ by which for a given rectangle the
condition \mref{A53} can be satisfied then none periodic orbit skeleton provides us with a possibility to construct on
it regular SWF's except the bouncing ball skeletons. For these two cases the energy spectra obtained
coincides with the one got from the "generic" skeleton calculations and their degeneracy disappears.

\subsection{Quantization of pseudointegrable systems - broken rectangles}

\hskip+2em By a broken rectangle we mean the one which can be decomposed into a finite set of disjoint rectangles, see
Fig.10. If reintegrated it shows some number of rectangular bays and peninsulas.
\begin{figure}
\begin{center}
\psfig{figure=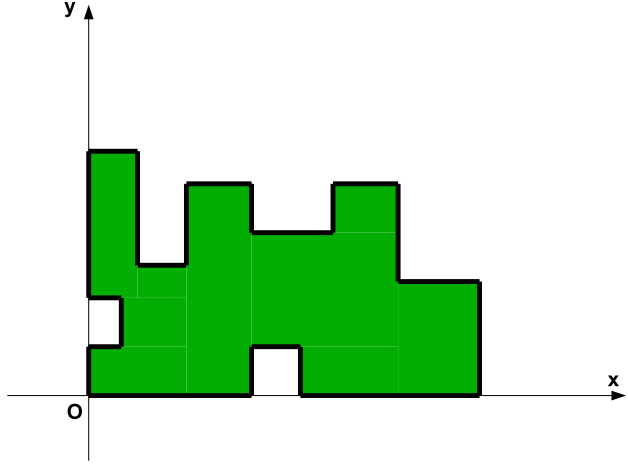,width=12cm}
\caption{An "arbitrary" broken rectangular billiards}
\end{center}
\end{figure}

In fact the broken rectangles can serve as archetypes of pseudointegrable systems with an arbitrary genus. Since however we are
interested in considering some special SWF's configurations related to classical periodic trajectories we shall limit
ourselves to rather simple forms of the broken rectangles. The simplest one with a single peninsula (the L-shaped
pseudointegrable billiards in terms of Kudrolli and Sridhar \cite{28}) is
shown in Fig.11 and also in Fig.12 and Fig.13 with several skeleton configurations related to some SD.

\begin{figure}
\begin{center}
\psfig{figure=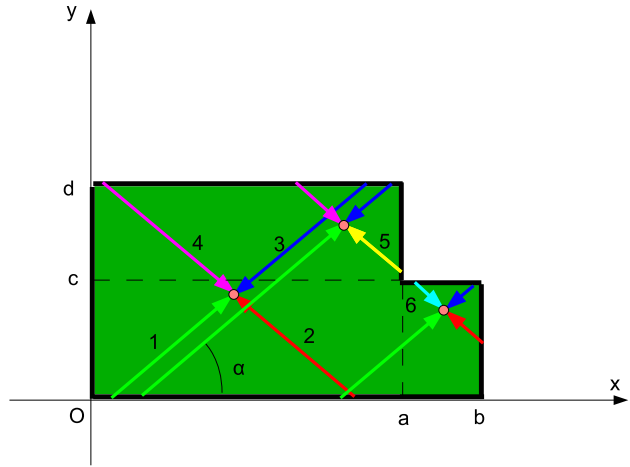,width=12cm}
\caption{A single bay rectangular billiards with a "generic" global singular skeleton composed of six bundles. Its
reduced form contains four global but singular compound bundles which form in the phase space the closed Lagrange
surface of genus 2.}
\end{center}
\end{figure}

\subsubsection{Regular SWF's in the broken rectangular billiards}

\hskip+2em Skeletons in the broken rectangular billiards which are to provide us with the SWF's which would be the exact solutions to the
corresponding eigenvalue problem have to be regular. Considering the billiards of Fig.11 it is clearly seen that there
are no such skeletons - each "generic" skeleton sketched
on Fig.11 has to have the bundles, denoted by 5. and 6. in the figure, which can be composed with the bundles 2. and 4.
respectively into two global but singular compound bundles. Therefore each "generic" skeleton has to be singular.
Despite this the Lagrange surface form by these skeletons are closed and of genus 2.

The first general conclusion which follows for the considered
case of the broken rectangular billiards and the more so for the more complicated ones
of Fig.10 is that one cannot expect the obtained SWF's to be exact.

To convince oneself of the correctness of the last conclusion we will consider the bouncing ball modes skeleton of Fig.12C,D, instead of
making a tedious calculations for the generic skeletons leading however to the same results. To this goal it is enough to match the corresponding
GSWF's defined on the skeletons shown in Fig.12C,D according to the conditions \mref{23a}. Both
the GSWF's have the form \mref{A36}. Therefore the respective procedure leads us to the following quantization conditions for the energy $E_{nm}$:

\begin{figure}
\begin{center}
\psfig{figure=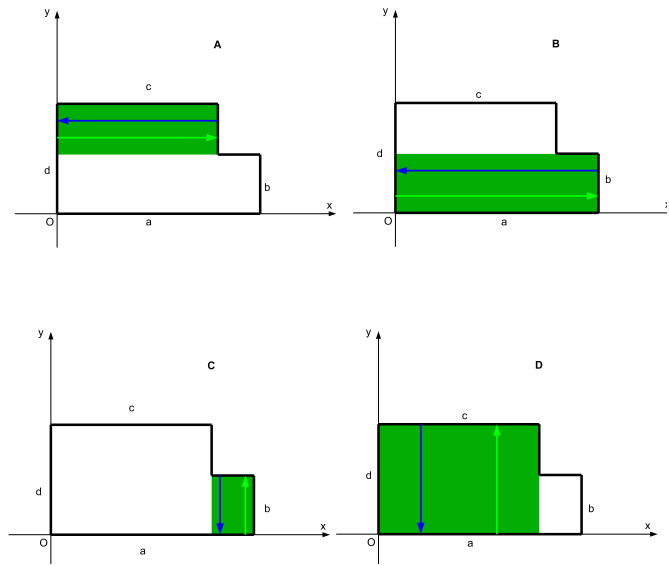,width=9cm}
\caption{A single bay rectangular billiards with singular skeletons built on periodic trajectories - the singular bouncing ball
cases. Every skeleton forms the cylinder-like Lagrange surface in the phase space.}
\end{center}
\end{figure}

\be
\frac{a}{n_0}=\frac{c}{l_0}=n\frac{\Lambda_{x}}{2},\;\;\;\;\;\;\;\;\frac{d}{m_0}=\frac{b}{k_0}=m\frac{\Lambda_{y}}{2}\nn\\
\Lambda_{x}=\frac{2\pi}{\sqrt{2E_1}},\;\;\;\;\;\;\;\;\Lambda_{y}=\frac{2\pi}{\lambda p}\nn\\
E_{nm}=\frac{2\pi^2}{\lambda^2}\ll(\frac{1}{\Lambda_{x}^2}+\frac{1}{\Lambda_{y}^2}\r)=
\frac{\pi^2}{2\lambda^2}\ll(\frac{n^2n_0^2}{a^2}+\frac{m^2m_0^2}{d^2}\r)\nn\\
m,n=1,2,...
\label{A58}
\ee
where $\Lambda_{x},\;\Lambda_{y}$ are the wave lengths of rays in the
horizontal and vertical skeletons respectively shown in Fig.12 and $n_0,m_0$ are the smallest integers satisfying $l_0a=n_0c$
and $k_0d=m_0b$ where $l_0,k_0$ are also integers.

The respective SWF's are the following:
\be
\Psi_{nm}^{as}(x,y,\lambda)=\ll\{\ba{lr}
A\sin\frac{2\pi x}{\Lambda_{x}}\sin\frac{2\pi y}{\Lambda_{y}}=
A\sin(\frac{nn_0}{a}\pi x)\sin\ll(\frac{mm_0}{d}y\r)&\;\;\;\;\;\;(x,y)\in D_{br}\\
0&\;\;\;\;\;\;(x,y)\notin D_{br}
\ea\r.
\label{A59}
\ee
where $D_{br}$ denotes the domain of the $x,y$-plane occupied by the broken rectangular of Fig.11.

The above GSWF is of course regular. Nevertheless due to the properties of the bouncing ball skeleton it is not exact despite the fact that it
satisfies the Dirichlet boundary conditions as well as the SE. The reasons for its approximate character are the conditions \mref{A58} which
cannot be satisfied if the corresponding length $a,b,c,d$ of the broken rectangle are not commensurate by pairs. However if such a commensurability
is satisfied by the sides of the broken rectangle then the solution \mref{A59} is exact.

One can also easily realize that the last results can be generalized to any bouncing ball modes skeleton in the broken rectangular
billiards of Fig.10. A little bit surprising is that the semiclassical formulae \mref{A58} for the energy and \mref{A59} for the wave functions
remain unchanged for any such a billiards while a number of conditions the wave lengths $\Lambda_{x}$ and $\Lambda_{y}$
have to satisfy filling the vertical and horizontal skeletons by integer numbers of their halves is increasing
respectively to numbers of bays and peninsulas forming the sides od such billiards. In fact for the corresponding SWF's the half wave lengths
$\fr\Lambda_{x}$ and $\fr\Lambda_{y}$ considered as the units of lengths on the respective horizontal and vertical sides of the broken rectangle
have to measure these sides by integers. It means of course that these sides have to be commensurate so theoretically such a condition excludes SWF's
for most the broken rectangular billiards. Practically however since incommensurability in fact does not exists in real measurements by
experimental errors one can always tune the corresponding waves to the real dimensions of the broken rectangles.

\begin{figure}
\begin{center}
\psfig{figure=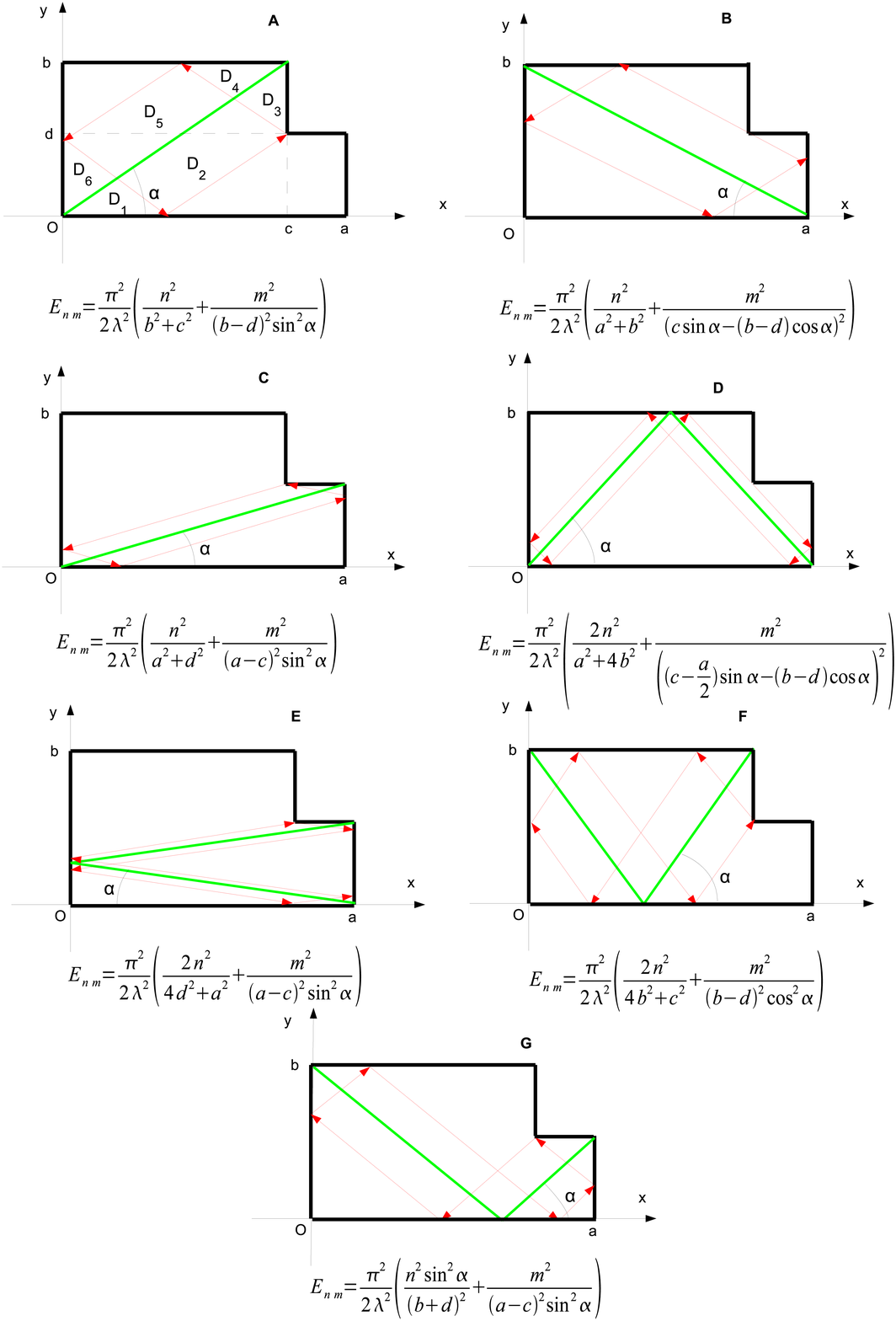,width=12cm}
\caption{A single bay rectangular billiards with singular skeletons built on the shortest periodic trajectories different
than the bouncing ball ones with the energy spectra of the respective GSWF's. Note that $n,m=1,2,...$, for
every spectrum. The red arrows show the outermost periodic orbits of the corresponding skeletons. The skeletons form the
cylinder-like Lagrange surface in the phase space each.}
\end{center}
\end{figure}

\subsubsection{Singular SWF's in the broken rectangular billiards}

\hskip+2em The SWF's \mref{A59} seem to be the unique regular ones which can be constructed in the broken rectangular billiards,
i.e. any other SWF's should be singular. Examples of them corresponds to all the skeletons shown in Fig.12 and Fig.13.
The skeletons of Fig.12 define singular SWF's which are identical with the ones of the formulae
\mref{A58} and \mref{A59} except that there are no relations between the rectangular sides $a,b,c,d$. These modes
were observed experimentally by Kudrolli and Sridhar \cite{28}.

Even more spectacular are skeletons built by periodic orbits different than the bouncing ball ones shown in Fig.13.
These are just the skeletons which provide us with the singular SWF's with properties described by Bogomolny and Schmit
\cite{1} as superscars and was observed also experimentally by Kudrolli and Sridhar \cite{28}.

The singular SWF's corresponding to the broken rectangle billiards shown in Fig.14 (upper figures)
were observed by Bogomolny {\it et al} \cite{2}. In fact the authors mentioned considered the limit of the billiards
when $d-c\to 0$ (lower figures). They studied experimentally
the high frequency modes in a microwave cavity \cite{2} confirming the existence of the superscar modes
predicted earlier by Bogomolny and Schmit \cite{1}.

\begin{figure}
\begin{center}
\psfig{figure=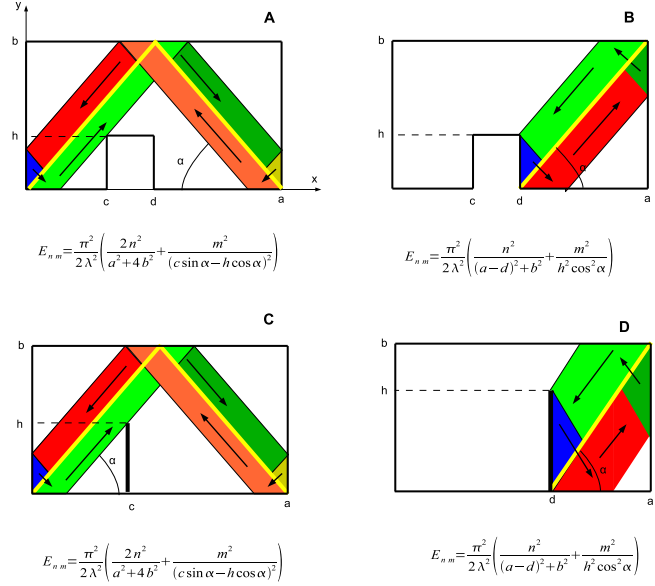,width=10cm}
\caption{A broken rectangular billiards with barriers and with possible superscar skeletons and with the energy spectra
corresponding to the respective GSWF's. In all the above formulae $n,m=1,2,...$. The corresponding Lagrange surfaces in
the phase space are cylinder-like.}
\end{center}
\end{figure}

For a completeness we shall give below the form of the singular SWF's for the skeleton $A$ of Fig.13 together with their
degenerate energy spectrum.
\be
\Psi_{1,2;mn}^{as;sing}(x,y)=\nn\\\ll\{\ba{l}
	        \sin\ll(\frac{m\pi}{c-d\cot\alpha}(x-y\cot\alpha)\r)\times
	        \ll\{\ba{l}\sin(\frac{n\pi}{D}(x\cos\alpha+y\sin\alpha))\\
	        \pm\cos(\frac{n\pi}{D}(x\cos\alpha+y\sin\alpha))
	        \ea\r.-\nn\\
	         \sin\ll(\frac{(m\pi}{c-d\cot\alpha}(x+y\cot\alpha)\r)\times
	          \ll\{\ba{l}\sin(\frac{n\pi}{D}(x\cos\alpha-y\sin\alpha))\\
	        \cos(\frac{n\pi}{D}(x\cos\alpha-y\sin\alpha))
	        \ea\r.\nn\\(x,y)\in D_1,\;D_6\nn\\
	        \sin\ll(\frac{m\pi}{c-d\cot\alpha}(x-y\cot\alpha)\r)\times
	        \ll\{\ba{l}\sin(\frac{n\pi}{D}(x\cos\alpha+y\sin\alpha))\\
	        \pm\cos(\frac{n\pi}{D}(x\cos\alpha+y\sin\alpha))
	        \ea\r.\nn\\(x,y)\in D_2,\;D_5\nn
	        \ea\r.\nn
	        \ee
	        \be\nn\\\ll\{\ba{l}
	        \sin\ll(\frac{m\pi}{c-d\cot\alpha}(x-y\cot\alpha)\r)\times
	        \ll\{\ba{l}\sin(\frac{n\pi}{D}(x\cos\alpha+y\sin\alpha))\\
	        \pm\cos(\frac{n\pi}{D}(x\cos\alpha+y\sin\alpha))
	        \ea\r.-\nn\\
	         \sin\ll(\frac{m\pi}{c-d\cot\alpha}(x+y\cot\alpha)\r)\times
	          \ll\{\ba{l}\sin(\frac{n\pi}{D}(x\cos\alpha-y\sin\alpha-2c\cos\alpha))\\
	        -\cos(\frac{n\pi}{D}(x\cos\alpha-y\sin\alpha-2c\cos\alpha))
	        \ea\r.\nn\\(x,y)\in D_3,\;D_4\nn
	        \ea\r.\nn\\
	        \tan\alpha=\frac{b}{c}\nn\\
\lambda pD=n\pi\nn\\
E_1=\frac{m^2\pi^2}{2(c\sin\alpha-d\cos\alpha)^2}\nn\\
E_{mn}=\fr p^2+\frac{E_1}{\lambda^2}=\frac{\pi^2}{2\lambda^2}\ll(\frac{n^2}{D^2}+\frac{m^2}{(c\sin\alpha-d\cos\alpha)^2}\r),\;\;\;\;\;\;m,n=1,2,3,...
\label{A60}
\ee
where $D=\sqrt{b^2+c^2}$ is the length of the diagonal shown in the Fig.13A.

\subsection{The rational polygon billiards other than the rectangular ones - the triangle and the pentagon
billiards}

\hspace{15pt}In this and in the next section we will made a short review of the billiards systems which have been
widely \cite{8,22,2,30,31} considered both theoretically and experimentally having mainly in mind their skeleton
description.

\subsubsection{The equilateral triangle billiards}

\hspace{15pt}These billiards which dynamics is integrable have been considered in past very often \cite{8,6}. From the skeleton
construction view point one can similarly to the rectangular case built the "generic" skeleton, see Fig.15a, as well as
skeleton generated by periodic orbits, Fig.15b-d.

\begin{figure}
\begin{center}
\psfig{figure=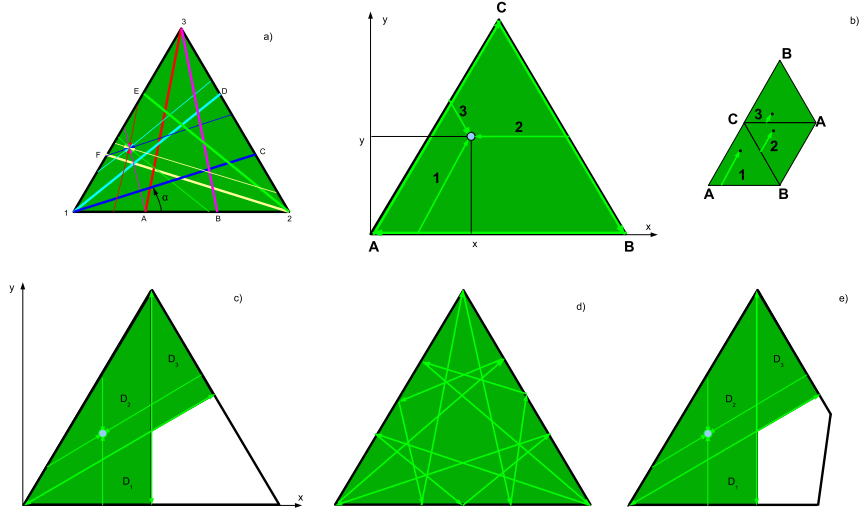,width=12cm}
\caption{Skeletons in the equilateral triangles: a) - a "generic" regular one forming a torus in the phase space, b)
- defined by periodic orbits (thick arrows) forming a M{\"o}bius band in the phase space, c) - defined by periodic orbits with a
cylinder-like Langrange surface, d) - with longer periodic orbits, e) - in the amputated corner triangle. The latter
can be neither integrable nor pseudointegrable.}
\end{center}
\end{figure}

In the "generic" case of Fig.15a the corresponding skeletons consist of twenty four bundles. The reduced skeletons
however have them already twelve and such a number counts a set of the BSWF's which as it is seen from Fig.15a have to
interfere to built the GSWF. The skeleton is regular and the result of such a superposition is well known \cite{8,6}
so we do not perform a corresponding calculations which leads to the exact eigenfunctions and the energy levels as
well of the corresponding eigenvalue problems.

It has to be stressed however that since the associated skeletons ${\bf B}^A$ coincide with {\bf B} then the energy
level degeneracy if happens has to be of different origins.

Nevertheless we will report here the corresponding results for the periodic skeletons of Fig.15b-c. The skeleton of
Fig.15b is defined by the SD composed of the three sides of the triangle so that the period of the orbit is equal
to three. The skeleton
contains only three bundles covering the whole triangle each so it is regular. To get GSWF it is enough to superpose in each point
of the triangle only three BSWF defined on the bundles. It is obvious also that for the skeleton shown in Fig.15b its
associated one does not coincide with it having the same energy, i.e. the corresponding energy spectrum is degenerate.
Taking therefore two independent superpositions of their GSWF we get the following regular semiclassical solutions for the case:
\be
\Psi_{mn}^{(1)}(x,y)=\sin\ll(\frac{2}{3}m\pi x\r)\sin\ll(\frac{2}{\sqrt{3}}n\pi y\r)-
\sin\ll[\frac{1}{3}m\pi(x+\sqrt{3}y)\r]\sin\ll[n\pi\ll(x-\frac{y}{\sqrt{3}}\r)\r]+\nn\\
\sin\ll[\frac{1}{3}m\pi(x-\sqrt{3}y)\r]\sin\ll[n\pi\ll(x+\frac{y}{\sqrt{3}}\r)\r]
\label{60}
\ee
and
\be
\Psi_{mn}^{(2)}(x,y)=\cos\ll(\frac{2}{3}m\pi x\r)\sin\ll(\frac{2}{\sqrt{3}}n\pi y\r)+
\cos\ll[\frac{1}{3}m\pi(x+\sqrt{3}y)\r]\sin\ll[n\pi\ll(x-\frac{y}{\sqrt{3}}\r)\r]-\nn\\
\cos\ll[\frac{1}{3}m\pi(x-\sqrt{3}y)\r]\sin\ll[n\pi\ll(x+\frac{y}{\sqrt{3}}\r)\r]
\label{60}
\ee
with the energy spectrum:
\be
E_{mn}=\frac{2\pi^2}{9\lambda^2}(m^2+3n^2)\nn\\
m,n=1,2,...
\label{62}
\ee
for both the solutions where $m,n$ are even or odd simultaneously.

Note that since both the above solutions are regular they are exact as well as their common energy spectra.

In the case of the periodic skeleton of Fig.15c defined by the periodic orbit of the length $\sqrt{3}$ (the double
height of the triangle) there are four bundles of the skeletons and the corresponding four BSWF's
which are singular since the skeleton is singular. Therefore the GSWF is defined locally in the domains
$D_k,\;k=1,2,3,$ of Fig.16c being however continuous in $D_1\cup D_2\cup D_3$. It is the following:
\be
\Psi_{mn}^{as}(x,y)=\nn\\\ll\{\ba{lr}
                   \sin\ll(2m\pi\frac{x}{\sqrt{3}}\r)\sin(2n\pi x)&(x,y)\in D_1\\
                   \sin\ll(2m\pi\frac{x}{\sqrt{3}}\r)\sin(2n\pi x)-
                   \sin\ll[m\pi\ll(x+\frac{y}{\sqrt{3}}\r)\r]\sin\ll[n\pi\ll(x+\sqrt{3}y\r)\r]&(x,y)\in D_2\\
                   -\sin\ll[m\pi\ll(x+\frac{y}{\sqrt{3}}\r)\r]\sin\ll[n\pi\ll(x+\sqrt{3}y\r)\r]&(x,y)\in D_3
                                      \ea\r.
\label{63}
\ee
with the energy spectrum:
\be
E_{mn}=\frac{2\pi^2}{3\lambda^2}(m^2+3n^2)\nn\\
m,n=1,2,...
\label{64a}
\ee

Of course $\Psi_{mn}^{as}(x,y)\equiv 0$ if $(x,y)\notin D_1\cup D_2\cup D_3$.

While the above $\Psi_{mn}^{as}(x,y)$ satisfies Schr{\"o}dinger equation its derivatives are not continuous on the boundaries separating
the domains $D_k,\;k=1,2,3$. This is why it is only semiclassical approximation to the exact solutions given earlier.

The spectrum \mref{64a} are naturally degenerate. However a source of this degeneracy is the symmetry of the equilateral
triangle. Namely, there are solutions with the spectrum \mref{64a} which can be obtained by rewriting the solution
\mref{63} in the new coordinate systems obtained from the present one by moving it to the remaining two vertices
and rotating it by $\pm\frac{2\pi}{3}$ respectively.

These new solutions can of course interfere with \mref{63} and with themselves so that trying to stimulate the
corresponding state in a cavity certainly such superposed states will be generated rather than the "pure" state
\mref{63}.

To isolate however the state \mref{63} it is enough to remove one of the three corners of the triangle as it is
shown in Fig.15e.

The skeleton corresponding to the periodic orbit shown in Fig.15d contains eighteen bundles none pair of which can be
done a compound one. The length of the orbit is equal to $D=3\sqrt{7}$ while the wideness of each bundle is equal to
$w=\fr\sqrt{\frac{3}{7}}$. Therefore the GSWF's for this case are singular and are linear combinations of the eighteen
BSWF's. The energy
spectrum is degenerate at least because the skeleton shown and its associated do not coincide. Its form can
be obtained using the following general formula for the energy spectrum of the periodic skeletons:
\be
E_{mn}=\fr p^2 +\frac{E_1}{\lambda^2}=\fr\frac{4m^2\pi^2}{\lambda^2D^2}+\frac{n^2\pi^2}{2\lambda^2w^2}=
\frac{\pi^2}{\lambda^2}\ll(\frac{2m^2}{D^2}+\frac{n^2}{2w^2}\r)
\label{64b}
\ee
where $D$ is the length of the orbit period and $w$ is the wideness of the skeleton stripe.

Therefore for the case considered we get:
\be
E_{mn}=\frac{2\pi^2}{63\lambda^2}(m^2+21n^2)\nn\\
m,n=1,2,...
\label{64}
\ee
with some possible relations between $m$ and $n$ which can follow from the corresponding possible relations between
$\chi$-factors of the BSWF's defined on the skeleton bundles and which can be established by detailed constructions of the
corresponding GSWF's.

\subsubsection{The pentagon billiards}

\hskip+2em  The pentagon form billiards were also the subject of intensive studies of both theoretical and experimental
\cite{30,31}. In the latter case the corresponding pentagon cavities were made of some dielectric media. In the very
high frequency region the corresponding electromagnetic waves form different modes among which the whispering gallery one
of Fig.16 was the most prominent. The other pentagon modes shown in the paper of Lebental {\it et al} \cite{30} are
more difficult for
an identification in terms of the corresponding skeletons also because of different boundary conditions the authors
wanted to consider.

\begin{figure}
\begin{center}
\psfig{figure=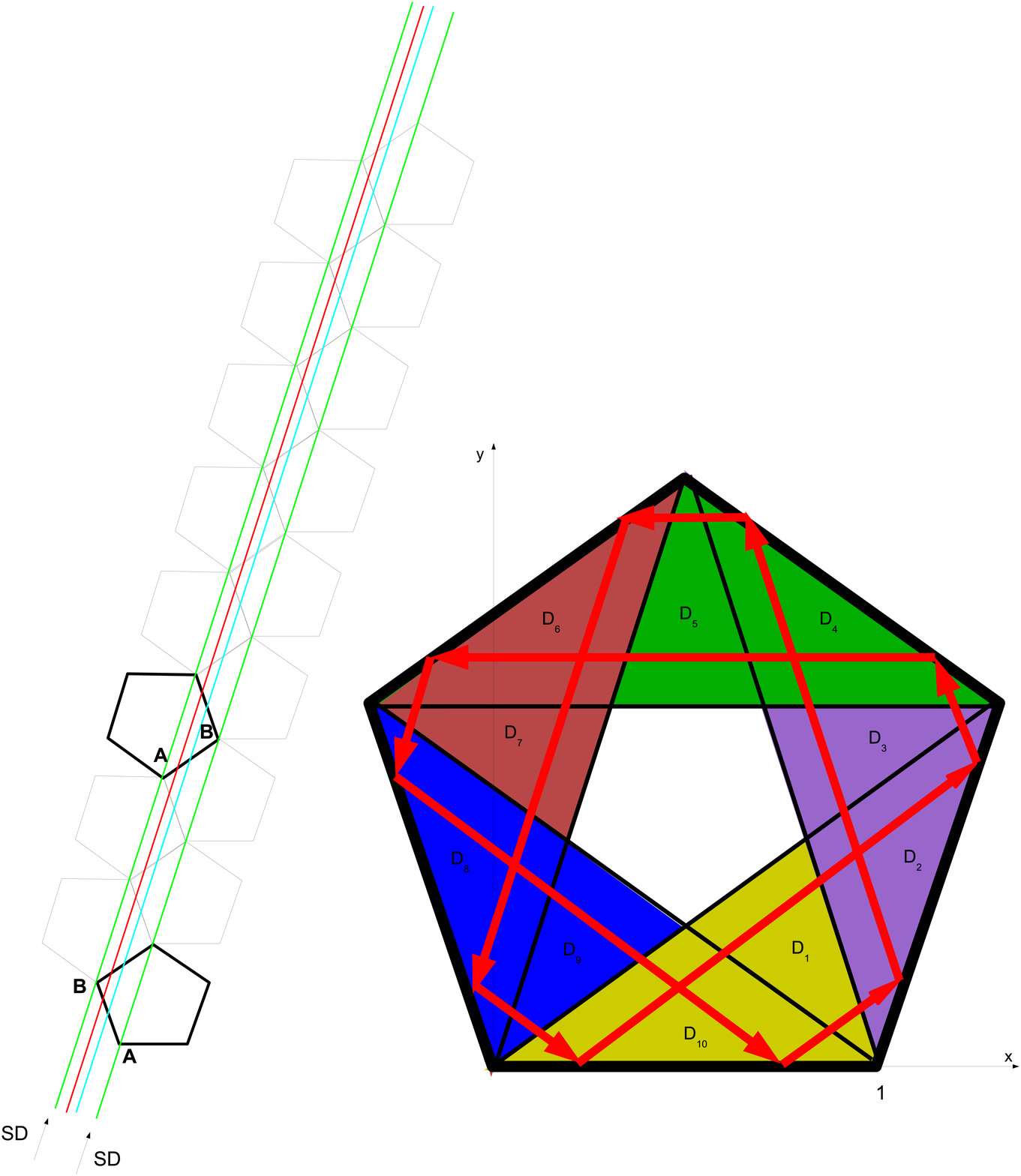,width=10cm}
\caption{The pentagon billiards and its simplest whispering gallery superscar skeleton with five bundles. A typical
periodic orbit is shown (the red straight line in the unfolded pentagon) the limiting form of which is the five-pointed star
orbit (the SD green straight lines). It degenerates into the inscribed pentagon orbit (the blue straight line) with the half of the
typical period equal to $10\cos\frac{\pi}{5}\simeq 8.090$. Gluing the end segments indicated by $AB$ in the unfolded pentagon
we get a M{\"o}bius band in the phase space.}
\end{center}
\end{figure}

Nevertheless in our paper we would like to distinguish other pentagon modes of SWF's. We will not however
consider in details a "generic" mode because of its complexity. Namely in such a case one can find simply by hand that there are
twenty compound bundles composing the corresponding skeletons and consequently the same number of BSWF's which have to interfere
to get the GSWF's satisfying the Dirichlet boundary conditions say. These compound bundles however are regular
each so the corresponding skeletons are regular and the GSWF's built on such skeletons will be regular and by that will
be exact solutions to the
corresponding SE with the exact energy spectrum despite the fact that these results can be obtained by the semiclassical approach.
Therefore instead of the "generic" skeleton cases we consider the singular ones generated by periodic orbits.

The simplest of such cases
is the whispering gallery skeleton shown in Fig.16. It is quite easy to write the corresponding singular GSWF. Nevertheless we limit ourselves to
quote merely the corresponding result for the energy spectrum. Namely we have:
\be
E_{mn}=\frac{\pi^2}{2\lambda^2}\ll(\frac{m^2}{25\cos^2\frac{\pi}{5}}+\frac{n^2}{\sin^2\frac{\pi}{5}}\r)\nn\\
m,n=1,2,...
\label{65}
\ee
where similarly to the triangle case the number $m,n$ are simultaneously even or odd.

The spectrum \mref{65} is of course degenerate. Note also that in the pentagon (white) center the corresponding GSWF's
vanish identically.

Another case of the singular skeleton shown in Fig.17 reminds the rectangular bouncing ball modes and the similar
triangle modes of Fig.15. The corresponding energy spectrum is:
\be
E_{mn}=\frac{\pi^2}{2\lambda^2}\ll(\frac{m^2}{(3\cos\frac{\pi}{10}+2\sin\frac{\pi}{5})^2}+\frac{n^2}{\sin^2\frac{\pi}{10}}\r)\nn\\
m,n=1,2,...
\label{66}
\ee

\begin{figure}
\begin{center}
\psfig{figure=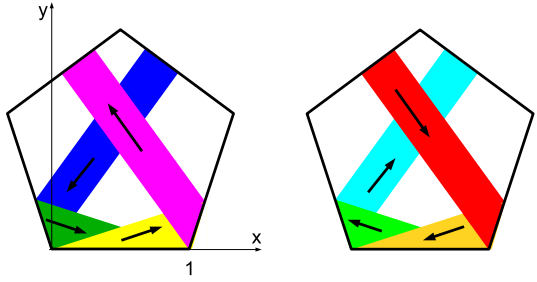,width=12cm}
\caption{Another simple singular periodic skeleton in the pentagon billiards with the half of the period equal to
$3\cos\frac{\pi}{10}+2\sin\frac{\pi}{5}\simeq 4.029$. Its Lagrange surface is cylinder-like.}
\end{center}
\end{figure}

\begin{figure}
\begin{center}
\psfig{figure=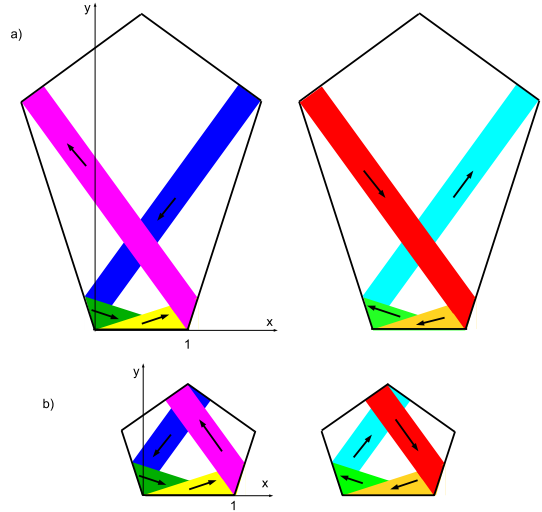,width=8cm}
\caption{The deformed pentagon billiards with the singular periodic skeletons having the half periods
equal to $2\cot\frac{\pi}{10}+\cos\frac{\pi}{10}\simeq 7.106$ - the case $a)$ and $2\tan\frac{\pi}{10}+3\cos\frac{\pi}{10}\simeq 3.503$
- the case $b)$}
\end{center}
\end{figure}

This spectrum is of course degenerate - there are five independent solutions with the same spectrum. In the pentagon cavity all these
solutions can be stimulated simultaneously. To isolate at least one of them it is enough to desymmetrize the pentagon into its forms
shown for examples in Fig.18. For the case $a)$ of the figure the corresponding energy spectrum is:
\be
E_{mn}=\frac{\pi^2}{2\lambda^2}\ll(\frac{m^2}{(2\cot\frac{\pi}{10}+\cos\frac{\pi}{10})^2}+\frac{n^2}{\sin^2\frac{\pi}{10}}\r)\nn\\
m,n=1,2,...
\label{67}
\ee
while for the case $b)$ it is:
\be
E_{mn}=\frac{\pi^2}{2\lambda^2}\ll(\frac{m^2}{(2\tan\frac{\pi}{10}+3\cos\frac{\pi}{10})^2}+\frac{n^2}{\sin^2\frac{\pi}{10}}\r)\nn\\
m,n=1,2,...
\label{68}
\ee

\section{Billiards with chaotic classical motions - superscars and periodic orbits}

\subsection{Singular SWF's in the chaotic polygon based billiards}

\hspace{15pt}It follows from the previous section that the idea of the skeletons seems to be effective in solving some
simple situations of quantum phenomena related semiclassically with billiards which shapes stimulate rather chaotic
than regular (integrable or pseudointegrable) motions. An example of such cases is
shown in Fig.15e. Still more spectacular situations exist in billiards which can be obtained from the rectangular and
the pentagonal ones
by their deformations. Examples of such deformations and the superscar modes possible to be detected in such
chaotic billiards are shown in Fig.18 and Fig.19. To describe analytically the superscar skeletons shown in these figures the methods
of the previous sections can be applied directly. Note that the superscar mode corresponding to the Sinai billiards of
Fig.19 was observed experimentally by Kudrolli and Sridhar \cite{28} and by Sridhar and Heller \cite{10} who
studied the Sinai billiards also numerically (see also \cite{16}).

\begin{figure}
\begin{center}
\psfig{figure=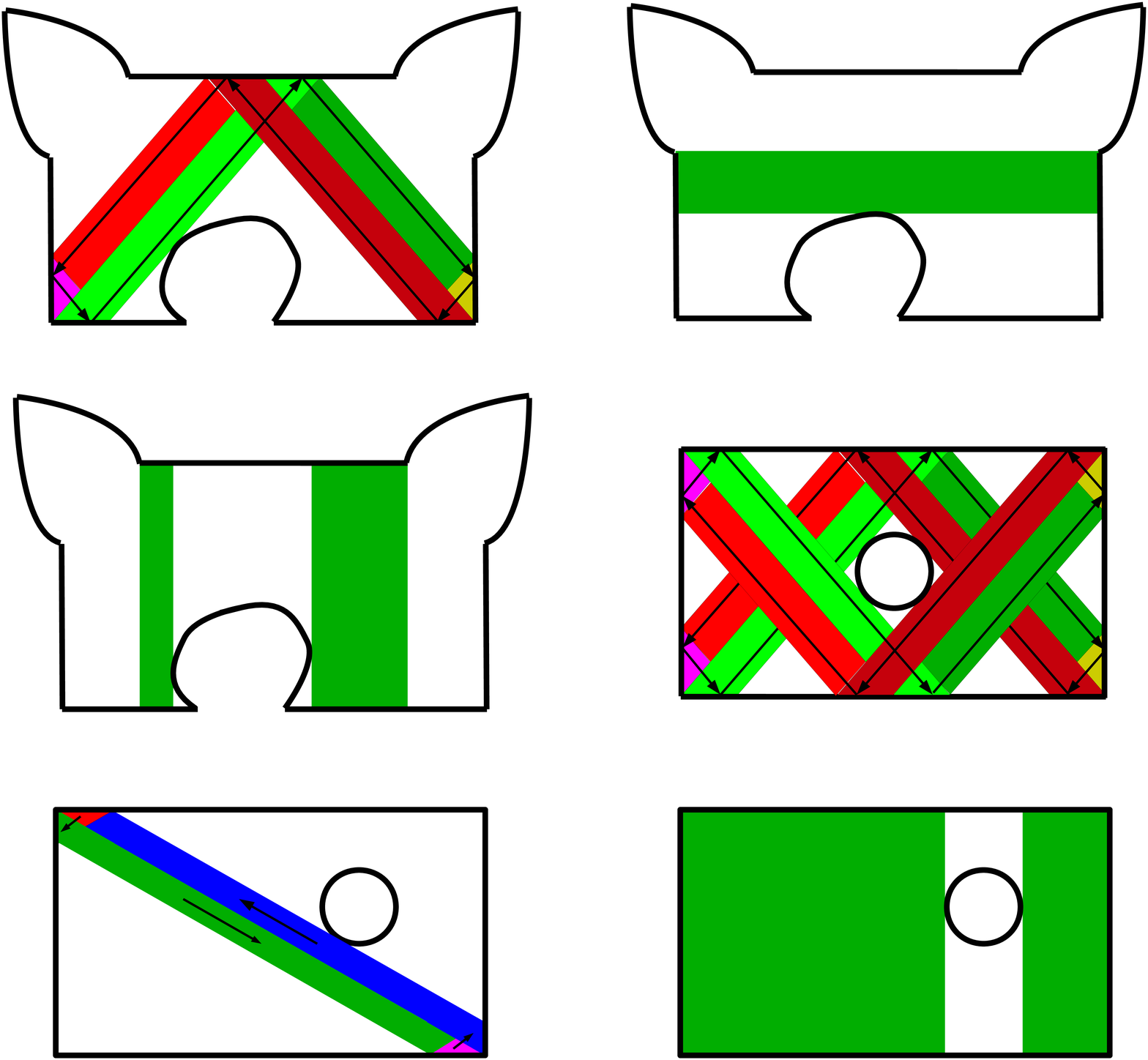,width=7cm}
\psfig{figure=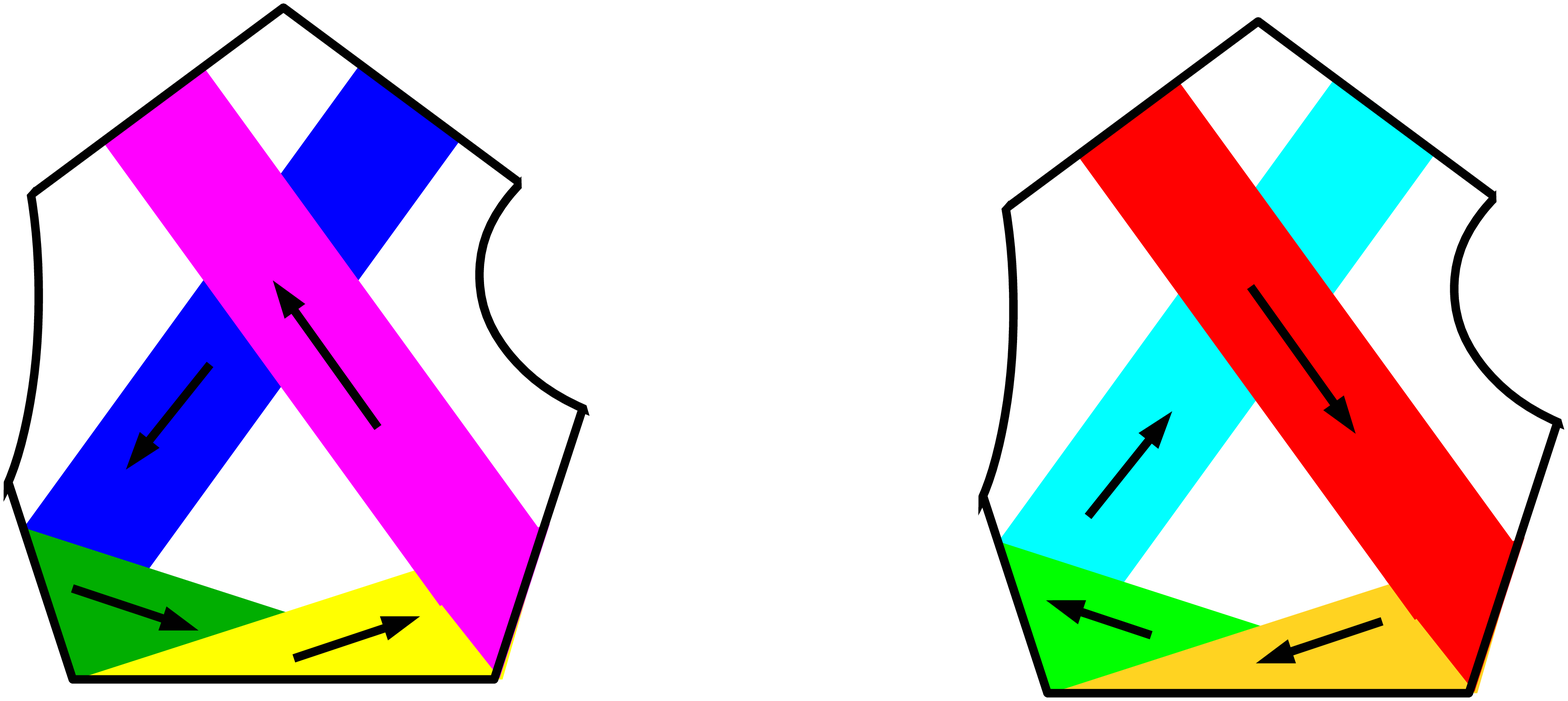,width=7cm}
\caption{The deformed rectangular, the Sinai and the pentagon billiards giving rise to chaotic motions with some
possible superscar skeletons}
\end{center}
\end{figure}

\section{Summary and conclusions}

\hspace{15pt}In this paper we have formulated very thoroughly the idea of skeletons in billiards and we have shown in great details
how to construct on them
semiclassical wave functions basing on the modified Maslov - Fedoriuk approach \cite{4}. The modification mentioned
utilizes rather the corresponding configuration spaces than the phasespaces making use of the complex time to continue
SWF's by caustics \cite{3}.

In general the skeletons in billiards play a role of the Arnold tori \cite{5} in the integrable
dynamical systems with smooth motions.

In this paper we have considered mainly the rational polygon billiards
beginning with the rectangular one to show that:
\begin{itemize}
\item a huge variety of skeletons can be found in such billiards;
\item by the Keller - Rubinov \cite{6} and the Ritchens - Berry \cite{8}
constructions global skeletons form in the phase space the closed Lagrangian surfaces of some genus ($\ge 1$);
\item the singular skeletons which are not global are represented in the phase space by open surfaces with
boundaries. Most of the periodic skeletons belong to such a class and form in the phase space the cylinder-like or
the M{\"o}bius-like bands;
\item the SWF's which solve the energy eigenvalue problems in such billiards can be of two
kinds - the regular and the singular ones - depending on the skeletons which they are constructed on;
\item the regular SWF's solve the energy eigenvalue problems exactly;
\item the singular SWF's solve the energy eigenvalue problems only approximately providing us with the superscar
solutions of Bogomolny and Schmit \cite{13,28,14,1,2,30,31};
\item the superscar skeletons and SWF's can be found in many chaotic billiards which contain flat boundaries in favourable
patterns \cite{28,14,16,10}.
\end{itemize}

Some general conclusion which can be done by reassuming our results is that in addition to the saturation of the Gutzwiller
formula \cite{32} for the semiclassical Green functions the other role of the periodic orbits is their being of a local organizer
of order in the integrable as well as in the pseudointegrable and the chaotic motions realized by the respective skeletons.

\section*{Appendix A}

\hspace{15pt} In this appendix we are going to show, that the geometrical optics rule of reflections of rays off the
billiards boundary is a consequence of demands of vanishing on the boundary of the linear combination \mref{19} accompanied by the conditions
\mref{19a} and \mref{20}. Namely consider the following superposition of SWF's:
\be
\Psi_k^{as}(x,y;u,l;\lambda)=\Psi_{k,1}^{\sigma_1}(d_1,s_1;u,l;\lambda)+\Psi_{k,2}^{\sigma_2} (d_2,s_2;u,l;\lambda)=\nn\\
J_{k,1}^{-\fr}(d_1,s_1;u,l)e^{i\sigma_1k\ll(d_1+\int_0^{s_1}\cos\alpha_{k,1}(s';u,l)ds'\r)}
\chi_{k,1}^{\sigma_1}(d_1,s_1;u,l;\lambda)+\nn\\
J_{k,2}^{-\fr}(d_2,s_2;u,l)e^{i\sigma_2k\ll(d_2+\int_0^{s_2}\cos\alpha_{k,2}(s';u,l)ds'\r)}
\chi_{k,2}^{\sigma_2}(d_2,s_2;u,l;\lambda)
\label{A19b}
\ee
with
\be
{\bf r}\equiv[x,y]={\bf r}_{k,1}(d_1,s_1;u,l)={\bf r}_0(s_1)+{\bf d}_{1}(s_1;u,l)=\nn\\{\bf r}_{k,2}(d_2,s_2;u,l)=
{\bf r}_0(s_2)+{\bf d}_{2}(s_2;u,l)\nn\\
{\bf r}_{k,1}(d,s;u,l)\in B_k(u,l),\;\;\;\;\;{\bf r}_{k,2}(d,s;u,l)\in B_k'(u,l),\;\;\;\;B_k(u,l)\neq B_k'(u,l)
\label{A19c}
\ee
i.e. the SWF's $\Psi_{k,1}^{\sigma_1} (d_1,s_1;u,l;\lambda)$ and $\Psi_{k,2}^{\sigma_2}(d_2,s_2;u,l;\lambda)$ are defined
respectively on the bundles $B_k(u,l)$ and $B_k'(u,l)$ with $D_k(u,l)\cap D_k'(u,l)\neq\oslash$ interfering in the
crossing point $[x,y]$ of two rays ${\bf r}_{k,1}(d_1,s_1;u,l)$ and ${\bf r}_{k,2}(d_2,s_2;u,l)$ belonging to the
respective bundles.

Therefore the condition for $\Psi_k^{as}(x,y,\lambda)$ to vanish on $A_k(u,l)$ is:
\be
J_{k,1}^{-\fr}(0,s)e^{ik\sigma_1\int_0^s\cos\alpha_{k,1}(s';u,l)ds'}\chi_{k,1}^{\sigma_1}(s;u,l;\lambda)+\nn\\
J_{k,2}^{-\fr}(0,s)e^{ik\sigma_2\int_0^s\cos\alpha_{k,2}(s';u,l)ds'}\chi_{k,2}^{\sigma_2}(s;u,l;\lambda)=0\nn\\
{\bf r}_0(s)\in A_k(u,l)
\label{A21}
\ee

Because of the $k$-dependence the last relation can be satisfied if and only if:
\be
\sigma_1\int_0^s\cos\alpha_{k,1}(s';u,l)ds'=\sigma_2\int_0^s\cos\alpha_{k,2}(s';u,l)ds',
\;\;\;\;\;\;{\bf r}_0(s)\in A_k(u,l)
\label{A21a}
\ee

It is easy to see however that there are only two solutions of the last condition:
\be
\alpha_{k,1}(s;u,l)\equiv\alpha_{k,2}(s;u,l)\;\;\;\;\;\;\;\; for\;\;\;\;\;\;\;\;\sigma_1=\sigma_2 \nn\\
\alpha_{k,1}(s;u,l)\equiv\pi-\alpha_{k,2}(s;u,l)\;\;\;\;\;\;\;\; for\;\;\;\;\;\;\;\; \sigma_1=-\sigma_2\nn\\
{\bf r}_0(s)\in A_k(u,l)
\label{A21c}
\ee

The first solutions are however uninteresting identifying the bundles in a given segment and
consequently leading to the solutions vanishing identically on $A_k(u,l)$.

Putting $\alpha_{k,1}(s;u,l)\equiv\alpha(s;u,l)$ and $\sigma_1=-\sigma$ we get from the second solution and from \mref{A21}:
\be
\chi_{k,2}^{-\sigma}(s;u,l;\lambda)=-\chi_{k,1}^{\sigma}(s;u,l;\lambda)\equiv\chi_k(s;u,l;\lambda)\nn\\
{\bf r}_0(s)\in A_k(u,l)
\label{A21b}
\ee
so that the combination \mref{19} becomes:
\be
\Psi_k^{as}(x,y;u,l;\lambda)=\Psi_{k,1}^{\sigma}(d_1,s_1;u,l;\lambda)-\Psi_{k,2}^{-\sigma} (d_2,s_2;u,l;\lambda)=\nn\\
J_{k,1}^{-\fr}(d_1,s_1;u,l)e^{i\sigma k\ll(d_1+\int_0^{s_1}\cos\alpha(s';u,l)ds'\r)}
\chi_{k,1}^{\sigma}(d_1,s_1;u,l;\lambda)-\nn\\
J_{k,2}^{-\fr}(d_2,s_2;u,l)e^{-i\sigma k\ll(d_2+\int_0^{s_2}\cos\alpha(s';u,l)ds'\r)}
\chi_{k,2}^{-\sigma}(d_2,s_2;u,l;\lambda)\nn\\
{\bf r}_0(s)\in A_k(u,l)
\label{A21d}
\ee
where $\chi_k{\sigma}(d,s;u,l;\lambda),\;\sigma=\pm,$ are given by \mref{13A} with
$\chi_k^{\sigma}(0,s;u,l;\lambda)\equiv\chi_k(s;u,l;\lambda)$.

The last result shows that $\Psi_k^{as}(x,y;u,l;\lambda)$ vanishing on $A_k(u,l)$ has to be
represented semiclassically by a
combination of at least two SWF's of opposite signatures and such that if $\Psi_{k,1}^{\sigma}(d,s;u,l;\lambda)$ is
defined on the bundle $B_k(u,l)$
then the second SWF $\Psi_{k,2}^{-\sigma}(d,s;u,l;\lambda)$ has to be defined on the bundle $B_k^A(u,l)$.

\section*{Appendix B}

\hspace{15pt}It is shown in this appendix that $\delta_k(u_j,l_j)$ from the formula \mref{22a} is
$s$-independent. To this end consider Fig.1  on which a mapping $h_k(s;u,l)$ of the arc $A_k(u,l)$ into an arc $A_{k'}(u',l')$ is defined
by the the bundle $B_k(u,l)$. According to this mapping we have:
\be
{\bf r}_0(h_k(s;u,l))={\bf r}_0(s)+{\bf D}(s;u,l)\nn\\
{\bf r}_0(s)\in A_k(u,l),\;\;\;\;\;\;{\bf r}_0(h_k(s;u,l))\in A_{k'}(u',l')
\label{C1}
\ee
where ${\bf D}(s;u,l)$ is a vector linking the point ${\bf r}_0(s)$ with the point ${\bf r}_0(h_k(s;u,l))$ of the billiards boundary.

Differentiating the equation \mref{C1} with respect to $s$ we get:
\be
\frac{\p h_k(s;u,l)}{\p s}\cos\beta(h_k(s;u,l))=\cos\beta(s)-D(s;u,l)\frac{\p\gamma_k(s;u,l)}{\p s}\sin\gamma_k(s;u,l)+\nn\\
\cos\gamma_k(s;u,l)\frac{\p D(s;u,l)}{\p s}\nn\\
\frac{\p h_k(s;u,l)}{\p s}\sin\beta(h_k(s;u,l))=\sin\beta(s)+D(s;u,l)\frac{\p\gamma_k(s;u,l)}{\p s}\cos\gamma_k(s;u,l)+\nn\\
\sin\gamma_k(s;u,l)\frac{\p D(s;u,l)}{\p s}
\label{C2}
\ee
where $ D(s;u,l)$ is the length of ${\bf D}(s;u,l)$.

Making further the proper linear combinations of the last equations we have finally:
\be
\frac{\p h_k(s;u,l)}{\p s}\cos\alpha_{k'}(h_k(s;u,l);u',l')=\cos\alpha_k(s;u,l)+\frac{\p D(s;u,l)}{\p s}\nn\\
-\frac{\p h_k(s;u,l)}{\p s}\sin\alpha_{k'}(h_k(s;u,l);u',l')=\sin\alpha_k(s;u,l)-D(s;u,l)\frac{\p\gamma_k(s;u,l)}{\p s}
\label{C3}
\ee
where we have taken into account the following relation between the angles involved:
\be
\alpha_{k'}(h_k(s;u,l);u',l')+\alpha_k(s;u,l)=\beta(h_k(s;u,l))-\beta(s)+2\pi\nn\\
\gamma_k(s;u,l)=\beta(s)+\alpha_k(s;u,l)
\label{C4}
\ee
which follows from Fig.1.

The independence of $s$ of $\delta_k(u_j,l_j)$ follows now easily from the first of the relations \mref{C3}.

\end{document}